\documentclass[12pt]{article}
\usepackage{amstex}
\usepackage{psfig}
\makeatletter
\newcommand{\mrm}[1]{\mathrm{#1}}
\renewcommand{\theequation}{\arabic{section}.\arabic{equation}}
\renewcommand{\b}{\mrm{b}}
\renewcommand{\c}{\mrm{c}}
\renewcommand{\d}{\mrm{d}}
\newcommand{\e}{\mrm{e}}
\newcommand{\g}{\mrm{g}}

\newcommand{\q}{\mrm{q}}

\newcommand{\Q}{\mrm{Q}}

\newcommand{\bbar}{\overline{\mrm{b}}}
\newcommand{\cbar}{\overline{\mrm{c}}}

\newcommand{\qbar}{\overline{\mrm{q}}}
\newcommand{\Qbar}{\overline{\mrm{Q}}}

\newcommand{\trv}{_{\perp}}

\newcommand{\Ha}[1]{{\rm H}_{#1}^{(1)}}
\newcommand{\BJ}[1]{{\rm J}_{#1}}
\newcommand{\BK}[1]{{\rm K}_{#1}}
\newcommand{\LQCD}{\Lambda_{\rm{QCD}}}
\newcommand{\mc}{m_{\c}}
\newcommand{\Mc}{M_{\c}}
\newcommand{\sla}{\hspace*{-0.20cm}/}
\newcommand{\Sla}{\hspace*{-0.22cm}/}
\newcommand{\jp}{J/\psi}
\newcommand{\lessim}{\raisebox{-0.8mm}%
{\hspace{1mm}$\stackrel{<}{\sim}$\hspace{1mm}}}
\newcommand{\alphas}{\alpha_{\mrm{s}}}
\newcommand{\alphaem}{\alpha_{\mrm{em}}}

\newcommand{\as}{{\mrm{AS}}}

\newcommand{\da}{{distribution amplitude}}
\newcommand{\das}{{distribution amplitudes}}

{\end{list}}
\newcounter{enumct}

\newlength{\abstwidth}
\begin{document}
 
\sloppy

\renewcommand{\arraystretch}{1.5}

\pagestyle{empty}

\begin{flushright}
CERN--TH/97--64  \\
WU B 97-3 \\
hep-ph/9704378
\end{flushright}
 
\vspace{\fill}
 
\begin{center}
{\LARGE\bf Higher Fock states and power counting}\\[3mm]
{\LARGE\bf in exclusive $\boldsymbol{P}$-wave quarkonium decays}\\[10mm]
{\Large Jan Bolz$^a$, Peter Kroll} \\[3mm]
{\it Fachbereich Physik, Universit\"at  Wuppertal}\\[1mm]
{\it D-42097 Wuppertal, Germany}\\[1mm]
{ E-mail: kroll@theorie.physik.uni-wuppertal.de}\\[2ex]
{\large and} \\[2ex]
{\Large Gerhard A. Schuler$^b$} \\[3mm]
{\it Theory Division, CERN,} \\[1mm]
{\it CH-1211 Geneva 23, Switzerland}\\[1mm]
{ E-mail: Gerhard.Schuler@cern.ch}
\end{center}
 
\vspace{\fill}
 
\begin{center}
{\bf Abstract}\\[2ex]
\begin{minipage}{\abstwidth}
Exclusive processes at large momentum transfer $Q$ factor into 
perturbatively calculable short-distance parts and long-distance 
hadronic wave functions. Usually, only contributions from the leading 
Fock states have to be included to leading order in $1/Q$. We show that 
for exclusive decays of $P$-wave quarkonia the contribution from 
the next-higher Fock state $|\Q\Qbar \g\rangle$ contributes at the same 
order in $1/Q$. We investigate how the constituent gluon attaches to the hard 
process in order to form colour-singlet final-state hadrons and argue that 
a single additional long-distance factor is sufficient to parametrize the 
size of its contribution. Incorporating transverse degrees of freedom and 
Sudakov factors, our results are perturbatively stable in the sense that
soft phase-space contributions are largely suppressed. 
Explicit calculations yield good agreement with data on $\chi_{\c J}$
decays into pairs of pions, kaons, and etas. We also comment on 
$\jp$ decays into two pions.
\end{minipage}
\end{center}

\vspace{\fill}
\noindent
\rule{60mm}{0.4mm}

\vspace{1mm} \noindent
${}^a$ Supported by Deutsche Forschungsgemeinschaft.\\
${}^b$ Heisenberg Fellow.

\vspace{10mm}\noindent
CERN--TH/97--64\\
April 1997

\clearpage
\pagestyle{plain}
\setcounter{page}{1} 
\section{Introduction}
%
Exclusive reactions at large momentum transfer $Q$ can be calculated 
in perturbative QCD (pQCD) owing to a factorization theorem \cite{Brodsky80}, 
which separates the short-distance physics of the partonic subreactions at 
the scale $Q$ from the longer-distance physics associated with the 
binding of the partons inside the hadrons. The full amplitude is given
as a sum of terms, where each term factors into two parts, 
a hard-scattering amplitude $T_{H}$, calculable in 
perturbative QCD, and wave functions $\psi(x_i,\bf{k}_{\perp i})$ for 
each hadron $H$. The importance of the various terms depends on their
scaling with $1/Q$. For the contribution with the weakest fall-off 
with $Q$ (leading-twist contribution), the amplitude $T_{H}$ describes 
the scattering of clusters of {\em collinear} partons from the hadron and 
is given by {\em valence-parton} scatterings only. Hence the only 
non-perturbative input required are the distribution amplitudes 
$\phi_H(x_i,Q)$ for finding valence quarks in the hadron, each 
carrying some fraction $x_i$ of the hadron's momentum. The \das\ 
represent wave functions integrated over transverse momentum ${\bf
k}\trv$ up to a factorization scale $\mu_F$ of order $Q$.  

Corrections to this standard hard-scattering approach (sHSA) 
(lowest-order in $\alpha_s(Q)$ calculations in the collinear approximation 
using valence Fock states only) can be of perturbative origin 
($\propto \alpha_s(Q)$) or power-like ($\propto 1/Q$). 
The latter may be classified as follows:
\begin{enumerate}
\item
  Corrections arising from the overlap of the soft wave functions.
\item
  Corrections associated with the transverse momentum of the partons
  inside the hadrons.
\item
  Corrections from higher Fock states.
\end{enumerate}
In order to show that predictions of the sHSA are reliable, two conditions 
have to be met. First, one has to prove that both the $O(\alpha_s(Q))$ 
corrections and the $1/Q$ corrections are small. 
Second, one has to ensure that the perturbative calculation of $T_H$ 
does not receive major contributions from phase space regions, 
where the virtualities of the internal partons become 
soft and perturbation theory is not applicable \cite{isg}. 

In recent years there has been progress in our understanding 
which of the above-mentioned corrections are important. 
On the one hand, the HSA has been modified through the incorporation
of tranverse degrees of freedom and gluonic radiative corrections in
form of a Sudakov factor \cite{BoLiSt}. Within this 
modified hard-scattering approach (mHSA), the  
perturbative contribution to exclusive observables can be calculated
self-consistently, because the Sudakov factor suppresses contributions 
from soft phase-space regions. In particular, the mHSA provides a
description of the pion--photon transition form factor, which is both
reliable and in agreement with data, already at momentum transfers as
low as $1\,$GeV \cite{JKR96,ref:Fpigam,mus:97}.

On the other hand, the overlap of the
initial- and final-state pion wave functions, representing a soft 
contribution  of higher-twist type, is still sizeable in the few GeV 
region in the calculation of the pion elastic form factor 
and just suffices to fill the gap between the 
perturbative prediction and the experimental data \cite{JaK93,JKR96}. 
Since the overlap
contribution, while commonly neglected, is an essential ingredient 
of the HSA \cite{Brodsky80}, the smallness of the perturbative 
contribution to the pion form factor should not be regarded as  
a failure but rather as consistent within the entire approach. 

In a recent paper \cite{BKS96} we have shown that the hierarchy of the 
$1/Q$ expansion for exclusive decays of heavy quarkonia is different 
from the canonical one: 
for $P$-wave states, the contribution from the next-higher Fock state 
is not suppressed by additional powers of $1/Q$ compared to the 
contribution from the leading Fock state. 
($Q$ is of the order of the heavy-quark mass $\mc$.) 
This is analogous to the case of {\em inclusive} decays of heavy quarkonia,
although there the long-distance matrix elements are organized into a 
hierarchy according to their scaling with $v$, the typical velocity of the 
heavy quark in the quarkonium \cite{Bod94}. 

For the $P$-wave charmonium state $\chi_{\c J}$, the Fock-state
expansion starts as
\begin{equation}
  |\chi_{\c J}(J^{++})\rangle = O(1)\, |\c\cbar_1({}^3P_J)\rangle
                      + O(v)\, |\c\cbar_8({}^3S_1)\, \g \rangle
                      + O(v^2) 
\ ,
\label{Fockexpansion}
\end{equation}
where the subscript $c$ specifies whether the $\c\cbar_c$ is in a 
colour-singlet ($c=1$) or colour-octet ($c=8$) state. 
The angular-momentum state of the $\c\cbar$ pair is denoted by the 
spectroscopic notation ${}^{2S+1}L_J$. 
Contrary to the case of $S$-wave decays, {\em two} $4$-fermion operators 
contribute to the decay rate of $P$-wave states 
into light hadrons at leading order in $v$
\begin{eqnarray}
 \Gamma[\chi_{\c J} \rightarrow \mathrm{LH}] & = & 
  \frac{c_1}{\mc^4}\, 
     \langle \chi_{\c J} | {\cal O}_1(^3P_J) | \chi_{\c J} \rangle 
\nonumber\\
  & &~ + \frac{c_8}{\mc^2}\, 
     \langle \chi_{\c J} | {\cal O}_8(^3S_1) | \chi_{\c J} \rangle 
  + O(v^2 \Gamma)
\ .
\label{eq:inclfact}
\end{eqnarray}
The term involving the colour-singlet matrix element is the one familiar 
from the quark-potential model. Indeed, up to corrections of order  $v^2$, 
one has 
$\langle \chi_{\c J} | {\cal O}_1(^3P_J) | \chi_{\c J} \rangle  = 
9 |R'_P(0)|^2 / (2\pi)$. 
The decays of $P$-wave (and higher orbital-angular-momentum states) 
probe components of the quarkonium wave function that involve 
dynamical gluons. However, the contribution to the inclusive 
annihilation rate from the higher Fock state $|\c\cbar \g \rangle$ is 
parametrized through a single number, namely the expectation value 
of the octet operator between the $\chi_{\c J}$ state, i.e.\ 
the colour-octet matrix element in (\ref{eq:inclfact}). 

Consider now the case of exclusive decays of $\chi_{\c J}$. For definiteness
take $\chi_{\c J} \rightarrow \pi \pi$ and define the amplitude 
$M_J$ through
\begin{equation}
  \Gamma[\chi_{\c J} \rightarrow \pi \pi ] = \frac{a_J}{\mc}\,
  \left| M_J \right|^2
\ ,
\end{equation}
where the $a_J$ are real numbers. The amplitude is the sum of a 
colour-singlet and a colour-octet part, $M_J = M_J^{(1)} + M_J^{(8)}$, 
corresponding to the leading and the subleading Fock state  
in (\ref{Fockexpansion}), respectively. 
The singlet amplitude factors into the hard amplitude 
$T_{HJ}^{(1)}$ describing the subprocess
\begin{equation}
  \c\cbar_1(^3P_J) \rightarrow \q\qbar'_1(^1S_0) + \q'\qbar_1(^1S_0)
\end{equation}
and the wave functions for the leading Fock states of the two pions 
and the $\chi_{\c J}$. The $\chi_{\c J}$ wave function can be taken in 
the non-relativistic limit, in which the charm (anticharm) momentum 
is given by $\frac{1}{2} p \pm K$ and terms linear in the relative 
momentum $K$ have to be kept. The only long-distance information required 
is hence the singlet decay constant $f_J^{(1)}$ or, equivalently, 
the derivative of the non-relativistic $\c\cbar$ wave function at the origin 
in coordinate space, $R'_P(0) \propto f_J^{(1)} \sqrt{\mc}$. 
The colour-singlet amplitude thus takes on the form
\begin{equation}
  M_J^{(1)} \sim \mc\, \alpha_s^2(\mc)\, 
     \left(  \frac{f_\pi}{\mc} \right)^2\, 
  \frac{f_J^{(1)}}{\mc^2}\, I_J^{(1)}
\ ,
\end{equation}
where $I_J^{(1)}$ is a convolution of the pion distribution 
amplitudes with a hard kernel, $I_J^{(1)}= 
\int \d x \d y \phi_\pi(x) \phi_\pi(y) f_J^{(1)}(x,y)$. 

For the calculation of the colour-octet amplitude two new ingredients 
enter, the octet wave function and the problem of colour conservation. 
Consider the wave function of the $|\c\cbar_8 \g\rangle$ state first. 
It is important to realize that in the $|\c\cbar_8\g\rangle$
Fock state not only the $\c\cbar$ pair is in a colour-octet state, but 
also the three particles, $\c$, $\cbar$ and $\g$, are in an $S$-state. 
Hence orbital angular momenta are not
involved and the transverse degrees of freedom can be integrated
over. Therefore, we only have to operate with a \da\ 
$\Phi_J^{(8)}(z_1,z_2,z_3)$ ($z_1 + z_2 + z_3 = 1$, where
$z_1$, $z_2$, $z_3$ are the plus light-cone fractions of $\c$, $\cbar$, 
and $\g$, respectively), that is, as usual, 
subject to the condition $\int \d z_1\d z_2 \Phi_J^{(8)}=1$, and 
an octet decay constant $f_J^{(8)}$ for each $J$. 
The colour-octet component of the $\chi_{\c J}$ is then given by 
\begin{equation}
  \label{coc}
  |\chi_{\c J}^{(8)} \rangle\, = \, \frac{t^a_{\c\cbar}}{2} \,
                      f_J^{(8)}\,\int \d z_1\d z_2
                      \Phi_J^{(8)}(z_1,z_2,z_3) \, S_{J\nu}^{(8)}
\ ,
\end{equation}
where $t=\lambda/2$ is the Gell-Mann colour matrix and $a$ the colour
of the gluon.  In the following, we will make 
the plausible assumption that the colour-octet $\chi_{\c J}$ states 
differ only by their spin wave functions, i.e.\ the \das\ as well as 
the decay constants are assumed to be the same for all $\chi_{\c J}$
states, $f_J^{(8)} = f^{(8)}$. 
The covariant spin wave functions in (\ref{coc}) are
readily constructed
\begin{equation}
  \label{csw8}
  S_{0\nu}^{(8)}\,=\, \frac{1}{\sqrt{6}}\,( p\sla\,+\, M_0)\, 
               ( p_{\nu}/M_0 -\gamma_{\nu} ), \quad
  S_{2\nu}^{(8)}\,=\, \frac{1}{\sqrt{2}}\,( p\sla\,+\,
  M_2)\,\varepsilon_{\mu\nu} \gamma^{\mu}
\ . 
\end{equation}

Owing to the non-relativistic expansion we can employ several
simplifications. First, the three partons of the $|\c\cbar\g\rangle$ 
Fock state can be taken to be collinear up to corrections of order $v$. 
Then, the $\c$ and $\cbar$ three-momenta scale as $\mc v$ and
differ by at most $\mc v^2$, i.e.\ $z_1 = z_2$ up to $O(v^2)$. 
And finally, the gluon momentum $|\vec{k}|$ 
is peaked at a value of the order of the binding energy $\epsilon 
= \Mc - 2 \mc$, where $\Mc$ is the average charmonium mass. 
Hence we can assume a $\delta$-function-like $\c\cbar\g$ distribution 
amplitude, $z_1=z_2=(1-z)/2$, $z_3=z$, where
$z \simeq \epsilon/\Mc \simeq 0.15$.
Therefore, analogously to the singlet case, the only long-distance 
information on the octet wave function that enters the final result is 
the octet decay constant $f^{(8)}$. 
Identifying the binding energy with $\mc v^2$ ($v^2 \sim 0.3$), we find 
$z_3 \sim v^2/2  \sim 0.15$,  
in accordance with the above estimate.\footnote{%
Since we are working to lowest order in the $v$-expansion we may simply 
take $p_{\c,\cbar} = z_{1,2} p$, $k=z_3 p$ 
in the actual calculation; the amount of off-shellness 
is $p_{\c}^2 - \mc^2 \sim \mc^2 v^2$ and $k^2 \sim \mc^2 v^4$.
Correspondingly, the gluon has three polarization states, cf.\ 
(\ref{csw8}).}

Next consider the problem of colour conservation. 
The obvious solution to colour conservation seems
to be to demand one of the final-state pions to be also in a higher
Fock state. The Fock-state expansion of the pion is governed by 
the ratio of the QCD scale and the hard scale $Q\sim \mc$ of the process, 
$\lambda = \LQCD /Q$. 
The (multipole-based) Fock-state expansion of the pion starts as follows
\begin{eqnarray}
  | \pi(0^{-+}) \rangle & = & O(1)\, |\q\qbar_1(^1S_0) \rangle
  + O(\lambda)\, |\q\qbar_8(^1P_1)\g \rangle
\nonumber\\
  & &~ + O(\lambda^2)\, \left\{ |\q\qbar_8(^3S_1)\g \rangle
  + \ldots \right\} + O(\lambda^3)
\ .
\label{Fockpion}
\end{eqnarray}
Both octet terms give corrections of the same order $\lambda^2$ with respect 
to the leading one: the extra suppression of the magnetic dipole transition
for ${}^3S_1$ ($\lambda^2$ rather than $\lambda$ as for an electric dipole
transition) is compensated by the suppression $\propto \lambda$ 
of $P$-wave production. The octet amplitude will then look as follows:
\begin{eqnarray}
  M_J^{(8)}[^1P_1] & \sim &  \mc\, \alpha_s^2(\mc)\, 
    \frac{f_\pi}{\mc}\,  \frac{f_\pi^{(P,8)}}{\mc^2}\,  
  \frac{f^{(8)}}{\mc^2}\, \frac{I_J^{(P,8)}}{\mc\lambda^2}
\nonumber\\
  M_J^{(8)}[^3S_1] & \sim &  \mc\, \alpha_s^2(\mc)\, 
    \frac{f_\pi}{\mc}\,  \frac{f_\pi^{(S,8)}}{\mc^3}\,  
  \frac{f^{(8)}}{\mc^2}\, \frac{I_J^{(S,8)}}{\lambda^2}
\ ,
\end{eqnarray}
where $I_J^{(8)}$ is a convolution of the hard process
\begin{equation}
 \c\cbar_8(^3S_1) \rightarrow \q\qbar'_8(^1P_1, {}^3S_1) + \q'\qbar_1(^1S_0)
\ ,
\end{equation}
the ordinary pion \da\ and the \da\  for the higher Fock states of the pion.
We have indicated the extra suppression of these higher Fock states by
assuming the octet decay constant to scale as 
$f_\pi^{(P,8)} \sim \lambda f_\pi$,  
$f_\pi^{(S,8)} \sim \lambda^2 f_\pi$,  and indicated explicitly the extra
suppression of $P$-wave production. The factor $\lambda^2$ ($\propto
1/m_c^2$) appears as a consequence of this particular solution of colour
conservation: the constituent gluon merely acts as a spectator which
runs from the $\chi_{\c J}$ to one of the pions without changing its momentum.

Yet, we disregard this possibility of colour conservation for the following 
two reasons. First, it requires the specification of two new \das\  
for the two higher Fock states of the pion. 
Second, the Fock-state expansion (\ref{Fockpion}) for the pion might be 
badly convergent. Even if a Fock-state expansion existed, it need not 
obey the usual multipole expansion assumed in (\ref{Fockpion}). 
If we do not want to work with explicit higher Fock components of 
the pion, we have to answer two questions: what the constituent 
gluon of the $|\c\cbar\g\rangle$ state couples to, and what determines 
the probability, i.e.\ what the analogue of 
the charmonium $f^{(8)}$ is on the pion side. 
To this end we invoke two assumptions. 

First, we take QCD perturbation theory to be valid down to 
virtualities of the order of $z \mc^2 \sim (\mc v)^2$ 
(see Figs.~4 and~5 and Appendix).
Then we can attach the gluon of the $|\c\cbar_8\g\rangle$ state 
(in all possible ways and with a coupling $\alpha_s^{\mrm{soft}}$) 
to the hard process leading to the Feynman diagrams shown in 
Figs.~\ref{fig:col8diags1} and \ref{fig:col8diags2}.  
Thus the hard process is 
\begin{equation}
 \c\cbar_8(^3S_1) \g \rightarrow \q\qbar'_1(^1S_0) + \q'\qbar_1(^1S_0)
\ .
\label{eq:octetreact}
\end{equation}

Second, the transverse motion of the valence quark and antiquark 
determines the importance of the momentum distribution of the constituent 
gluon. It is clear that
the diagrams corresponding to (\ref{eq:octetreact}) must contain 
contributions, which, to leading order in $\alpha_s$ and $z$, constitute 
the two higher Fock states $|\q\qbar\g\rangle$ of the pion. 
In the approximation of collinear light quarks, one thus encounters 
singular integrals. These infrared singularities precisely correspond  
to the long-distance wave functions describing the higher Fock states
of the pion. In a collinear calculation one hence needs \das\ for the higher
Fock states into which the singularities can be absorbed. 
In our ansatz we keep the (non-perturbative) transverse motion of the 
pion's valence constituents (quark and antiquark). Then all integrals 
are finite. More precisely, gluon momentum configurations are selected 
such that the $\q$ and $\qbar$ relative momentum ``matches'' the one 
prescribed by the pion wave function. In this sense one can say that 
our approach leads to the dynamical generation of higher Fock components. 
We emphasize that also in this scheme of colour conservation the 
colour-octet contribution is not suppressed by powers of $1/m_c$ as 
compared to the colour-singlet one. This has been discussed by us in 
\cite{BKS96}, see also Sect.~4.

In our previous calculation \cite{BKS96}, we employed the collinear 
approximation and simply regularized the singular  
integrals through a cut-off, which corresponds to an average
transverse momentum of the quarks inside the pions. 
The final results for the colour-octet contribution to the decay amplitude 
then basically depended on a single parameter 
\begin{equation}
  \label{oldkappa}
  \kappa = \sqrt{\alphas^{\rm soft}} f^{(8)}/\varrho^2
\ .
\end{equation}
Here $\alpha_s^{\rm soft}$ denotes the coupling of the gluon of the 
$|\c\cbar \g\rangle$ Fock state to the hard process, and 
$f^{(8)}$ is the octet decay constant. 
There is an additional contribution from terms not singular 
for $\varrho \rightarrow 0$, which is
proportional to $\sqrt{\alphas^{\rm soft}} f^{(8)}$. It is, however, 
rather small, less than about 10\%. 

In this work we are going to extend our previous one by working within 
the mHSA. This leads to three major improvements. 
First, the scale of the coupling constant is not fixed 
in the collinear approximation, implying a big uncertainty. 
In contrast, the scales of $\alpha_s$ are fixed in the mHSA. 
Second, in the collinear approximation there is usually no suppression 
of the end-point regions of the \da, while these infrared regions 
are explicitly suppressed in the mHSA through a Sudakov form factor. 
And finally, no ad-hoc regularization of singular integrals 
appearing in the calculation of the octet contribution is necessary. 
Rather, the momentum distribution of the octet gluon is linked to 
the $k_{\perp}$ dependence of the pion wave function. 

The paper is organized as follows: In Sect.~2 we present
our ansatz for the pion wave function and in Sect.~3 we introduce 
the mHSA in the case of the colour-singlet decay contribution
to $\chi_{\c J} \to \pi^+ \pi^-$. In Sect.~4 and Appendix, a
detailed description of the colour-octet contribution and a discussion
of the results for the decay widths are given. In Sect.~5 we
present results for the $\chi_{\c J}$ decays into pairs of kaons and
etas. Finally, in Sect.~6 we comment on $\jp$ decays, 
before we end our paper with a summary and some conclusions (Sect.~7).
%

\section{The pion wave function} 
%
\setcounter{equation}{0}
We write the pion's $| \q\qbar\ \rangle$ Fock state, which is the
leading one, in a covariant fashion 
\begin{equation}
  \label{pion}
  |\,\pi \,\rangle = \frac{\delta_{ij}}{\sqrt{3}} \int 
       \frac{\d x\,\d^2 {\bf k}_{\perp}}{16\pi^3} \,
       \Psi_{\pi}(x,k_{\perp}) \, S_{\pi} 
\end{equation}
where the spin wave function is defined as (the pion mass is neglected
throughout)
\begin{equation}
  \label{spion}
  S_{\pi} = \frac{1}{\sqrt{2}} p\sla \gamma_5 \:.
\end{equation}
Integrating the light-cone wave function $\Psi_{\pi}$ over
$k_{\perp}$, the transverse momentum of the quark with respect to the
pion momentum, up to a factorization scale $\mu_F$, one arrives, 
up to a constant dimensionful factor $f_{\pi}/(2\,\sqrt{6})$, at 
the \da\ $\phi_\pi(x,\mu_F)$. The constant factor plays the role 
of the configuration space wave function at the origin; 
$f_\pi = 130.7\,$MeV is the usual pion decay constant. 

Upon expansion over Gegenbauer polynomials $C_n^{(3/2)}$, the \da\  
at a scale $\mu_F$ can be characterized by non-perturbative 
coefficients $B_n$ (see \cite{Brodsky80} and references therein):
\begin{equation}
  \label{phiex}
  \phi_\pi(x,\mu_F) = \phi_{\as}(x)\, 
  \left[ 1 + \sum_{n=2,4,\ldots}^{\infty}\, B_n(\mu_0)
  \left( \frac{\alphas(\mu_F) }{\alphas(\mu_0) } \right)^{\gamma_n}
  \, C_n^{(3/2)}(2x-1) \right] \, .
\end{equation}
The odd expansion terms do not appear since, in the isotopic  
limit, the pion \da\ is symmetric $\phi_{\pi}(x) = \phi_{\pi}(1-x)$. 
In (\ref{phiex}) $\alphas$ is the strong coupling constant, and 
$\mu_0$ a typical hadronic scale, $0.5 \lessim \mu_0 \lessim 1\,$GeV.
Since the anomalous dimensions $\gamma_n$ in (\ref{phiex}) are
positive fractional numbers increasing with $n$, higher-order terms
are gradually suppressed and any \da\ evolves into $\phi_{\as}(x) =
6x(1-x)$ asymptotically, i.e.\ for $\ln(Q^2/\LQCD^2) \rightarrow
\infty$. The asymptotic (AS) distribution amplitude itself shows no
evolution. 

From the investigation of the pion--photon transition form factor
$F_{\gamma \pi}(Q^2)$ \cite{JKR96,ref:Fpigam,mus:97,rad:95,Rau96} 
it follows that the form of the pion \da\ is very close to the 
asymptotic form. Also recent QCD sum-rule analyses \cite{bra:94} 
lead to that result. Hence, all terms in the expansion (\ref{phiex}) 
of the pion \da\ with $n \ge 2$ will provide only small corrections to
exclusive observables and it is legitimate to truncate that expansion
at the second term (note that the frequently used Chernyak--Zhitnitsky
\da\  \cite{CZ84} is given by $B_2=2/3$ and $B_n=0$
for $n\geq 4$) and consider now $B_2$ as the only soft parameter. 

\begin{figure}
 \[
   \psfig{figure=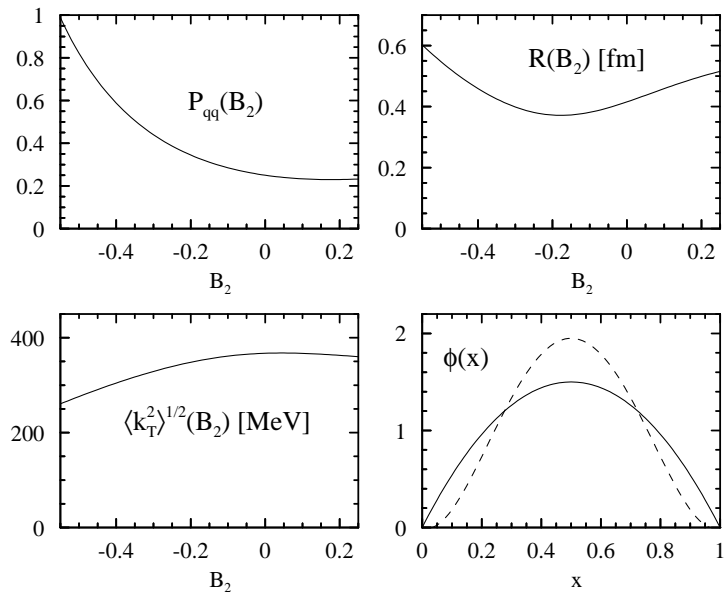,%
            bbllx=15pt,bblly=290pt,bburx=505pt,bbury=690pt,%
            clip=,height=8cm}
 \]
\caption[]{$B_2$ dependence of the probability $P_{\q\qbar }$ 
(upper left), the radius $R$ (upper right), r.m.s.\ transverse momentum 
$\langle k_{\perp}^2 \rangle^{1/2}$ (lower left) of the pion's 
valence Fock state as well as the shape of the \da\  (lower right). 
In the lower right figure, the solid (dashed) line
corresponds to $B_2 = 0$ ($-0.2$).}
\label{fig:pionwf}
\end{figure}

In the mHSA we consider the Fourier transform of $\Psi_{\pi}$ to
transverse coordinate space which, following \cite{JKR96,JaK93}, is written as
\begin{equation}
  \label{Psihat}
  \hat \Psi_{\pi}(x,{\bf b},\mu_F)  =  \frac{f_{\pi}}{2\,\sqrt{6}} \, 
  \phi_{\pi}(x,\mu_F) \, \hat\Sigma_{\pi}(x,b,\mu_F) \,.
\end{equation}
The dependence on the transverse separation
${\bf b}$, canonically conjugated to ${\bf k}_{\perp}$, is thereby 
chosen to be of a simple Gaussian form
\begin{equation}
   \label{Sigma}
  \hat \Sigma_{\pi}(x,b,\mu_F)   =   4\pi \, 
        \exp\left[-\frac{x (1-x)\,b^2}{4\,a_{\pi}^2(\mu_F)} \right] \,.
\end{equation}  
Here, the momentum fraction $x$ and ${\bf b}$ (or ${\bf k}_{\perp}$)  
refer to the quark; the antiquark momentum is
characterized by $1-x$ and ${\bf b}$ (or ${\bf k}_{\perp}$) throughout. 
The pion's transverse size parameter $a_{\pi} = a_\pi(\mu_F)$ is fixed
from the process $\pi^0 \rightarrow \gamma\gamma$ \cite{BHL83}. That 
constraint leads to the closed formula 
$1/a_\pi^2 =  8\, (1+B_2(\mu_F))\,\pi^2\,f_{\pi}^2$ 
under the assumption $B_n=0$ for $n\geq 4$ (for the
asymptotic \da\ ($B_2=0$) $a_{\pi}=0.861$ GeV$^{-1}$ \cite{JKR96}).
The scale dependence of $a_{\pi}$ is an approximation sufficient for
our purpose.
Using (\ref{pion})--(\ref{Sigma}) we obtain the following
expressions for the probability of finding the pion in its valence Fock
state, for the mean transverse momentum and the mean radius of the 
$\q \qbar$
Fock state  
\begin{eqnarray}
     \label{pionprops}
  P_{\q \qbar} & = & \frac{1}{4}\; \frac{1 + 18/7 B_2^2}{1 + B_2}
                                                           \nonumber \\
  \langle k\trv^2 \rangle & = & \frac{4\pi^2}{5}\,f_{\pi}^2\, (1 + B_2) 
      \frac{1 - 6/7\,B_2 + 12/7\,B_2^2}{1 + 18/7\,B_2^2} 
                                                            \nonumber\\
  R_{\q \qbar}^2 & = & \frac{3}{8\pi^2}\, f_{\pi}^{-2} \,
     \frac{1 + 3 B_2 + 54/7 B_2^2}{(1 + B_2)^2} 
\end{eqnarray}
As shown in Fig.\ \ref{fig:pionwf}, our wave function has appealing
features as a function of $B_2$. For the asymptotic form of the \da\ 
we obtain the values $P_{\q \qbar} = 0.25$, 
$\langle k\trv^2 \rangle^{1/2} = 367$ MeV and $R_{\q \qbar} = 0.42$ fm. 
We remark that the latter is considerably smaller than the pion's
charge radius of 0.66 fm \cite{amend} to which all the Fock states contribute.
Since $P_{\q\qbar }$ is much smaller than unity, higher Fock states are
important components of the pion and, of course, have been seen for instance in
Drell--Yan dilepton ($\mu^+ \mu^-$) production in pion--proton collisions
\cite{sut92}. At large Bjorken $x$, where the valence quarks dominate,
our wave function, for small $B_2$, is consistent with the valence
quark distribution function as derived in \cite{sut92} (see \cite{JKR96}).
For large negative values of $B_2$, i.e.\ at a very low scale, the
wave function bears resemblance to a constituent wave function with a
probability close to unity and a valence Fock state radius that
is approximately equal to the charge radius of the pion.

In exclusive reactions, contributions from higher Fock states of the pion are
usually neglected since they are suppressed by inverse powers of the
relevant hard scale, see (\ref{Fockpion}). As explained in the introduction, 
we do not make use of explicit Fock-state wave functions of the pion. 
Rather we generate the contribution corresponding to a constituent gluon 
of the pion dynamically. This is achieved in two steps:  
(i) the constituent gluon of the $\chi_{\c J}$ is coupled 
perturbatively ($\propto \alpha_s^{\mrm{soft}}$) to the hard process; 
(ii) the dependence of the momenta of the pion's valence constituents
(i.e.\ the $\q$, $\qbar$ momenta) on $k_{\perp}$ is kept.
Since it is the non-zero value of $k_{\perp}$ that regularizes the 
propagators associated with the ``would-be'' constituent gluon of the pion, 
the $k_{\perp}$ dependence of the wave function of the $\q\qbar$ valence 
Fock state determines the size of the $\q\qbar \g$ contribution in the pion. 
%
\section{The colour singlet contribution and the mHSA
\label{sect:HSP}}
%
\setcounter{equation}{0}
In \cite{BKS96} we have analysed the colour singlet contribution
to $\chi_{\c J} \to \pi^+ \pi^-$ already within the mHSA. 
Here, we will give a more detailed description of the calculation. 
  
The colour singlet components of the $\chi_{\c J}$ are expressed in
terms of non-relativistic wave functions 
\begin{eqnarray}
  \label{csin}
  |\,\chi_0^{(1)}\,\rangle\,&=& \, \frac{\delta_{ij}}{\sqrt{3}}\,
                      \int\frac{\d^3k}{16\pi^3 M_0}
                      \tilde{\Psi}_0^{(1)}(k)\, S_0^{(1)}, \nonumber \\
  |\,\chi_2^{(1)}\,\rangle\,&=& \, \frac{\delta_{ij}}{\sqrt{3}}\,
                      \int\frac{\d^3k}{16\pi^3 M_2}
                      \tilde{\Psi}_2^{(1)}(k)\, S_2^{(1)} 
\ , 
\end{eqnarray}
where the $\tilde{\Psi}_J^{(1)}$ represent reduced wave functions,
i.e.\ full wave functions  with a factor of the relative momentum 
$K_{\mu}$ removed from them ($Kp =0$ and, for instance, 
$K_{\mu}=(0,{\bf k})$ in the meson rest frame). 
The $S_J^{(1)}$ denote the covariant spin wave functions 
\begin{eqnarray}
  \label{csw0}
  S_0^{(1)}\,&=&\, \frac{1}{\sqrt{2}} \left [p\sla\,+\, M_0\,+\,
    2K\Sla \right ]\, K\Sla, \nonumber\\
  S_2^{(1)}\,&=&\,\frac{1}{\sqrt{2}} \left [( p\sla\,+\, M_2)\,
                 \gamma_{\rho}\,+\,\frac{2}{M_2}\left [( p\sla\,+\,
              M_2)\,K_{\rho}\, - p\sla K\Sla \gamma_{\rho}\right ]\right ]
                              \varepsilon ^{\rho\sigma} K_{\sigma}.
\end{eqnarray}
The spin wave functions represent an expansion upon powers of $K$
up to terms of $O(K^2)$ \cite{hus91} \footnote{
The spin wave functions can be written in a more compact form:
$S^{(1)}_J = \tilde{S}^{(1)}_J + \{\tilde{S}^{(1)}_J,K\Sla \}/M_J$
where $\tilde{S}^{(1)}_J$ represents the $O(K)$ spin wave function.}. 
The derivative of the non-relativistic $\c \cbar$ wave 
function at the origin is introduced by 
\begin{equation}
  \label{dcc}
  R'_P(0)\, =\, \imath\,\frac{16}{3}\,\pi^{3/2}\,\sqrt{\mc}\, 
          \int \frac {\d k k^4}{16\pi^3 M_0} \tilde{\Psi}_0^{(1)} \, 
                     \, =\, \imath\sqrt{16\pi \mc} f^{(1)}_0
\ .
\end{equation}
In the non-relativistic approximation one has 
$M_0\, \simeq\, M_2\, \simeq \,2\mc$
and the same normalization of the two wave functions (usually unity)
$P_0^{\c\cbar}\, =\, P_2^{\c\cbar}$. In this case one finds
$\tilde{\Psi}_2\, =\, \sqrt{3}\tilde{\Psi}_0$ and the same relation 
between the two singlet decay constants $f^{(1)}_J$.

The starting point of the calculation of the colour singlet decay 
amplitude within the modified HSA is the convolution with 
respect to the momentum fractions $x,y$ and transverse separation 
scales ${\bf b}_1,{\bf b}_2$ of the two pions 
\begin{multline}
  M^{(1)}(\chi_{\c J} \to \pi^+\pi^-) = -\imath\,
      \frac{32\,\sqrt{2}\,\pi^{3/2}\:|R'_P(0)|}
      {3\sqrt{3}\,\mc^{7/2}} \sigma_J\, \,
      \int_0^1 \d x \d y \, \int \frac{\d^2 {\bf b}_1}{(4\pi)^2}
                                 \frac{\d^2 {\bf b}_2}{(4\pi)^2} \\
      \hat \Psi_{\pi}^{\ast}(y,{\bf b}_2)\,
      \hat T_{HJ}(x,y,{\bf b}_1,{\bf b}_2)\,
      \hat \Psi_{\pi}(x,{\bf b}_1)\,
      \exp[-S(x,y,{\bf b}_1,{\bf b}_2,t_1,t_2)]
\ ,
  \label{Mb}
\end{multline}
which adapts the method proposed by \cite{BoLiSt} to our case of exclusive
charmonium decays ($\sigma_0=1, \sigma_2=\sqrt{3}/2$). $\hat T_{HJ}$ is the
Fourier transform of the hard-scattering amplitude $T_{HJ}$ with the
$k\trv$-dependence retained 
\begin{equation}
   \label{fourier}
   \hat T_{HJ}(x,y,{\bf b}_1,{\bf b}_2) = 
       \int \frac{\d^2 {\bf k}_{\perp 1}}{(2\pi)^2} 
            \frac{\d^2 {\bf k}_{\perp 2}}{(2\pi)^2} \:
       T_{HJ}({\bf k}_{\perp 1},{\bf k}_{\perp 2})\,
       \exp\left[- \imath\,{\bf k}_{\perp 1} {\bf \cdot b}_1
                - \imath\,{\bf k}_{\perp 2} {\bf \cdot b}_2 \right] \,.
\end{equation}  
\begin{figure}
\[
 \psfig{figure=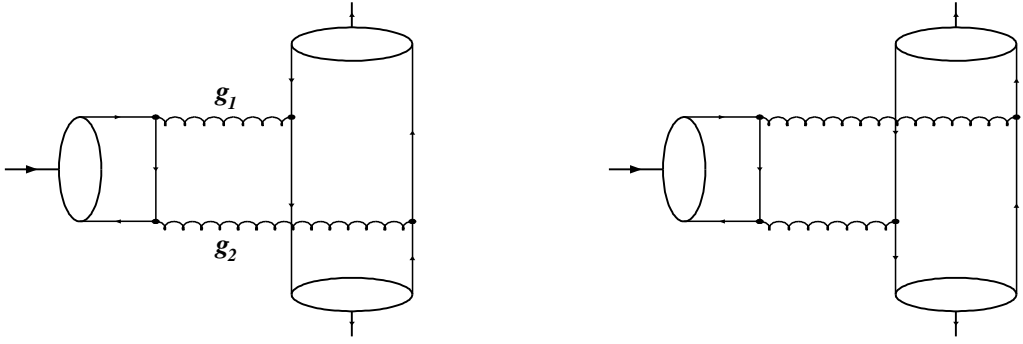,%
       bbllx=50pt,bblly=40pt,bburx=750pt,bbury=285pt,%
       height=5cm,clip=} \]
 \caption[dummy0]{Feynman graphs for the colour-singlet decay
  $\chi_{\c J} \rightarrow \pi\, \pi$ ($J=0,2$).  
  \label{fig:col1graph} }
\end{figure}
$T_{HJ}$ is to be calculated from the graphs shown in
Fig.~\ref{fig:col1graph} and reads
\begin{equation}
  {T}_{HJ}(x,y,{\bf k}_{\perp 1},{\bf k}_{\perp 2}) 
    = \frac{\alphas(t_1)\,\alphas(t_2)}
           {(\tilde{g}_1^2 + \imath\epsilon)\,(\tilde{g}_2^2 
                                         + \imath\epsilon)\,N} \,
      \left( 1 + \frac{(-2)^{J/2}}{2}\frac{(x-y)^2}{N} \right) 
\ .
  \label{THkt}
\end{equation}
With 
\begin{equation}
      \label{kperp}
{\bf K}_{\perp} = {\bf k}_{\perp 1}\!-\!{\bf k}_{\perp 2}
\end{equation}
the virtualities of the internal quark and the two gluons are
\begin{eqnarray}
  N & = &  x (1-y) + (1-x) y + \,{\bf K}^2_{\perp}/(2\mc^2) \, ,\nonumber \\
  \tilde {g}_1^2 & = &  x y - {\bf K}^2_{\perp}/(4\mc^2) \, ,\nonumber \\
  \tilde {g}_2^2 & = & (1-x)(1-y) - {\bf K}^2_{\perp}/(4\mc^2) 
\ , 
  \label{Ngi}
\end{eqnarray}
where a common factor $4\mc^2$ is pulled out.
The convolution formula (\ref{Mb}) involves two
independent transverse separation scales $b_1$ and $b_2$, each
associated with one of the two pions. The $b_i$ also provide the
factorization scales $\mu_{Fi} = 1 / b_i$ ($i=1,2$) of the pion wave 
functions. Distinct from these scales are the renormalization scales 
$t_j$ ($j=1,2$) corresponding to the virtualities of the hard gluons. 

The fact that $T_{H J}$ depends on ${\bf k}_{\perp 1}$ and ${\bf
k}_{\perp 2}$ only in the combination ${\bf K}_{\perp}$ implies the 
following result for the Fourier transform of the hard-scattering
amplitude ($\tilde b \equiv 2 \mc b_1$)
\begin{eqnarray}
  \hat T_{HJ}(x,y,{\bf b}_1,{\bf b}_2) 
                    & = & 4\,\mc^2\,\alphas(t_1^2) \alphas(t_2^2) 
                                \delta^{(2)}({\bf b}_1 - {\bf b}_2)
        \bigg\{ \frac{-\imath/4}{(1-x) (1-y) - x y} \, \nonumber\\
  & \phantom{-}&      \bigg[ 
           \frac{\Ha{0}(\sqrt{x y} \, \tilde b)}{x+y}
               \left(1 + \frac{(-2)^{j/2}(x-y)^2}{2 (x+y)}
               \right) \nonumber \\
  & - & 
        \frac{\Ha{0}(\sqrt{(1-x)(1- y)} \, \tilde b)}{2-x-y}
               \left(1 + \frac{(-2)^{j/2}(x-y)^2}{2 (2-x-y)} \right) 
        \bigg] \nonumber \\
  & + & \frac{1}{\pi} 
        \frac{\BK{0}(\sqrt{(x (1-y) + (1-x) y)/2}\,\tilde b)} 
             {(x+y)(2-x-y)}\,\left(1 + \frac{(-2)^{j/2}(x-y)^2}
             {(x +y)(2-x-y)} \right) \nonumber \\
  & + & \frac{1}{8 \pi} \frac{(-2)^{j/2}(x-y)^2}{(x+y)(2-x-y)}  
        \, \frac{\tilde b \, 
                 \BK{1}(\sqrt{(x (1-y) + (1-x) y)/2}\,\tilde b)}
                {\sqrt{(x(1- y) + (1-x) y)/2}}
             \bigg\}\: 
  \label{THb}
\end{eqnarray}
$\Ha{0}$ and $\BK{i}$ 
denote Hankel and modified Bessel functions, respectively.
In physical terms the appearance of the $\delta$-function, which
simplifies the numerical work enormously, means that the two pions 
emerge from the decay with identical transverse separations. 

The novel ingredient of the mHSA is the Sudakov factor
$\exp[-S]$, which takes into account those gluonic radiative
corrections not accounted for in the QCD evolution of the wave
function. In next-to-leading-log approximation the Sudakov exponent reads 
\begin{multline}
  S(x,y,{\bf b}_1,{\bf b}_2,t_1,t_2) =  
           s(x,b_1,2 \mc) + s(1-x,b_1,2 \mc) \\
         + s(y,b_2,2 \mc) + s(1-y,b_2,2 \mc) 
         - \frac{4}{\beta} \log
               \frac{\log(t_1/\LQCD)\log(t_2/\LQCD)}
                    {\log(1/(b_1\LQCD))\log(1/(b_2\LQCD))} \,,
  \label{sudexp}
\end{multline}
where the function $s(x,b,Q)$, originally derived by Botts and Sterman
\cite{BoLiSt} and later on slightly improved, can be found, for
instance, in \cite{DJK95}. The last term in (\ref{sudexp}) arises 
from a renormalization group transformation from the factorization 
scales $\mu_{F i}$ to the renormalization scales $t_j$ at which the 
hard amplitude $\hat T_H$ is evaluated. The renormalization scales 
$t_j$, entering $\hat T_H$ as the arguments of the two strong coupling
constants $\alphas$, are chosen as
\begin{equation}
  t_1 = \max\{4 x y\, \mc^2,1/b_1^2,1/b_2^2\}\, ,   \hspace{1cm}
  t_2 = \max\{4 (1-x) (1-y)\, \mc^2,1/b^2_1,1/b^2_2\} \, ,
  \label{tj}
\end{equation}
thus avoiding large logs from higher-order pQCD. In the limit $b_1
\to 1/\LQCD$ the $t_j$ may in principle approach $\LQCD$ at which
the one-loop expression for $\alphas$ is diverging. Still the integral
appearing in (\ref{Mb}) is regular since the Sudakov factor
compensates the $\alphas$ singularities. Actually the suppression of
the end-point regions is so strong that the bulk of the perturbative
contribution is accumulated in regions of small values of $\alphas$.
The physical picture behind the Sudakov suppression is that $\q\qbar $
pairs with large mutual separation tend to radiate so many gluons that
it becomes impossible for the hadronic state to remain intact and that
the exclusive process can take place. 
 
Replacing $\exp[-S]$ by 1 and ignoring the transverse momenta in
$T_{H}$, one finds from (\ref{Mb}) the colour singlet decay amplitude
within the standard HSA as derived by Duncan and Mueller
\cite{dun:80}. In that approach the renormalization scale is taken as
the charm quark mass and customarily identified with the
factorization scale.

In terms of the amplitude (\ref{Mb}) the $\chi_{\c J}$ decay widths
into pions are given by 
\begin{eqnarray}
  \label{width}
    \Gamma[\chi_{\c 0} \to \pi^+ \pi^-] & = & \frac{1}{32\,\pi\,\mc}\:
    \left| M(\chi_{\c 0} \to \pi^+ \pi^-)\right|^2 \label {Gamma0} \nonumber\\
  \Gamma[\chi_{\c 2} \to \pi^+ \pi^-] & = & \frac{1}{240\,\pi\,\mc}\:
    \left| M(\chi_{\c 2} \to \pi^+ \pi^-)\right|^2 
\ .
\label {Gamma2} 
\end{eqnarray}
The results obtained within the mHSA can be cast into the form 
\begin{equation}
  \label{decayzero}
  \Gamma[ \chi_{\c J} \rightarrow \pi^+\pi^-] = 
      \frac{f_{\pi}^4}{\mc^8}\, |R'_P(0)|^2\, \alphas(\mc)^4\, 
  \left| a_J^{(1)} + b_J^{(1)}\, B_2(\mu_0) 
       + c_J^{(1)}\, B_2(\mu_0)^2 \right|^2 \ , 
\end{equation}
where the coefficients $a_J$, $b_J$, and $c_J$ are complex-valued.
Written in the form (\ref{decayzero}) we may then
immediately find the prediction for a \da\ with a given value of 
the expansion coefficient $B_2$ at the scale $\mu_0$. It is still 
convenient to divide out the fourth power of $\alphas$ at the fixed 
scale $\mc$ in (\ref{decayzero}), since the main effect of the strong 
coupling is thus made explicit. The actual effective renormalization 
scale $\mu_R$ (i.e.\ the mean value of $\mu_R$) in the modified HSA 
differs from $\mc$ and depends on $B_2$ (e.g.\ $\mu^{eff}_R = 1.15$ 
GeV and hence $\alphas = 0.43$ for $B_2(\mu_0)=0$).

In the following we will choose as central value $R'_P(0)= 0.22$
GeV$^{5/2}$ and $\mc = 1.5\,$GeV, which is consistent with a global 
fit of charmonium parameters \cite{MaP95} as well as results for 
charmonium radii from potential models \cite{BuT81}.  
Moreover, we use the one-loop expression for $\alpha_s$ with 
$\LQCD = 200\,$MeV and four light flavours. 
As we have demonstrated in \cite{BKS96} by varying the input parameters 
$\mc$ and $R'_P(0)$, $B_2$, and $\LQCD$, the colour-singlet 
contribution alone, i.e.\ a calculation based on the assumption that
the $\chi_{\c J}$ is a pure $\c\cbar$ state, is insufficient to explain
the observed decay rates. 
%
\section{The colour octet contribution \label{IV}}
%
\setcounter{equation}{0}
Recently, the importance of higher Fock states in understanding the 
production and the {\em inclusive} decays of charmonia has been pointed 
out \cite{Bod94}. The heavy-quark mass allows for a systematic expansion 
of both the quarkonium state and the hard, short-distance process and, 
hence, of the inclusive decay rate or the production cross section. The
expansion parameter is provided by the velocity $v$ of the heavy quark
inside the meson. 
The crucial observation is that, for inclusive decays 
(and also production rates) of the $\chi_{\c J}$, 
both states in (\ref{Fockexpansion}) contribute
at the same order in $v$ and, hence, the inclusion of the 
``octet mechanism'', i.e.\ the contribution from the 
$|\c\cbar_8\g\rangle$ state, is necessary for a consistent description.
(Without its inclusion the factorization of the decay width into long-
and short-distance factors is spoiled by the presence of infrared-sensitive
logarithms.) The inclusion of the octet mechanism is also necessary in
exclusive charmonium decays in order to get a consistent
description. This is so because the usual suppression of higher Fock
states in exclusive reactions does not appear for the decay of
$P$-wave charmonium: it is compensated for by the $P$-wave nature of
the $\c\cbar_1$ Fock state \cite{BKS96}.

As a consequence of employing the collinear 
approximation we encountered in \cite{BKS96} a number of singular integrals, 
which we regularized by a parameter $\varrho$ related to the
(neglected) transverse momentum of the internal quarks and
gluons. However, within the modified HSA where the 
dependence on the partonic transverse momenta is taken into account, all
integrals are finite and there is no longer any need for the parameter
$\varrho$. Apart from that, the modified HSA also
provides an explicit prescription for the renormalization scales
entering the strong coupling constant such that $\alphas$ is no longer
a free parameter. Altogether, the modified HSA not only guarantees a
self-consistent and thus reliable calculation of the perturbative
contribution, it also helps in the present case to diminish
uncertainties of the collinear approximation. 

As explained in the introduction, the Fock-state gluon has a typical 
momentum fraction $z \sim \mc v^2/2 \mc$  of the $\chi_{\c J}$ momentum. 
We treat $z$ as a small number. We have checked that the results 
are nearly $z$-independent for $0.1 < z < 0.3$. Hence the $|\c\cbar\g\rangle$ 
wave function $\Phi_J^{(8)}$ of (\ref{coc}) can be taken as exhibiting 
a $\delta$-function-like peak at the value $z_3 = z$. Moreover, in our
results we will neglect all terms of order $z^3$ and higher. 
This is in line with the non-relativistic expansion of the charmonium 
decay rate since $z$ scales as $v^2$. 
Apart from $z$, the only free
parameter on the charmonium side is $f^{(8)}$, the colour octet wave
function at the origin. 
 
The details of the calculation of the colour octet contribution within
the modified HSA can be found in the Appendix. We have
divided the diagrams contributing to the colour octet decay into eleven
groups, numbered by the index $i$. The contribution of each group is
of the form (\ref{Mb}). Furthermore, we have introduced
renormalization scales $t_{ij}$, which are determined by the
virtualities of the internal quark and gluon lines (see Appendix) 
and depend non-trivially on the integration variables. As
already discussed in Sect.~\ref{sect:HSP} we may simply evaluate the
running coupling constant by its one-loop expression at the scales
$t_{ij}$ because its singularity (to be reached for $b_1$ or $b_2 \to
\LQCD$) is compensated by the Sudakov factor.   
  
We now want to direct the reader's attention to the question of
gauge invariance. Treating the problems as presented above, the results 
we obtain are gauge invariant to order $z^2$ (with $z$ being the
fraction of the $\chi_{\c J}$ momentum carried by the valence
gluon), which is sufficient, because we neglect terms of $O (z^3)$
in our results. Eventual violations of gauge-invariance at order $z^3$
may be traced back to the general problem to describe a constituent
within the parton model. This conjecture is supported by the following
observation we made: if one changes the $\c$-quark propagator mass in
groups  8--10 from the usual value $\mc$ to $(1\!-\!z)\mc$, we end up
with completely gauge-invariant results.  
Admittedly, this choice is by no means mandatory and we do not apply
it actually, but it reveals that the violations of gauge invariance to
$O (z^3)$ are related to the neglect of binding effects for which that
particular choice of the mass in the c-quark propagators seems to
compensate for.

In order to combine colour singlet and octet contribution we expand the
colour-octet decay amplitude (\ref{Moct}) in a fashion analogous to the
colour singlet case: 
\begin{equation}
  \frac{1}{\sqrt{\lambda_J \pi \mc}}\frac{1}{|\,R_p'(0)\,|} 
       \left(\frac{\mc^4}{\alphas(\mc)\,f_{\pi}}\right)^2 \,
       M^{(8)}(\chi_{\c J} \to \pi^+ \pi^-)  =    
                             \, \left( 
       a^{(8)}_J + b_J^{(8)} B_2 + c_J^{(8)} B_2^2 \right) 
\ ,
  \label{Octetcoeffs}
\end{equation} 
where $\lambda_0 = 32$ and $\lambda_2 = 240$. 
Thus, to obtain the sum of
singlet and octet contributions, we simply have to add the
coefficients, e.g.~$a_J^{(1+8)} = a^{(1)}_J + a^{(8)}_J$. 

First, we determine a value for the unknown decay constant $f^{(8)}$, 
which is contained in the coefficients $a^{(8)}_J$,
$b_J^{(8)}$ and $c_J^{(8)}$, 
from a fit of our predictions to the experimental data on the decay
widths, using the asymptotic pion wave function ($B_2=0$). We obtain
\begin{equation}
   f^{(8)} = 1.46 \times 10^{-3} \,\mbox{GeV}^2 \,,
   \label{foctet}
\end{equation}    
which is larger than the result estimated in
Ref.~\cite{BKS96}. This is mainly a consequence of the treatment of
the strong coupling constant $\alphas$. In Ref.~\cite{BKS96} we have chosen
0.45 for the coupling of the two hard gluons and assumed
$\alphas^{\rm soft} \approx \pi$ for the coupling of the soft
$\chi_{\c J}$ Fock-state gluon to the hard process: $\alphas(t_1)
\alphas(t_2) \sqrt{\alphas(t_3)} = 0.45^2 \sqrt{\pi} \approx 0.36$. 
In the present analysis based on the mHSA we may extract
effective values for the strong coupling constant. We distinguish the
hard $\alphas$, evaluated at the scales $t_{i1}$ and $t_{i2}$ from the
soft one at $t_{i3}$. Using the asymptotic form of the pion \da\ we
obtain   
\begin{equation}
 (\alphas^{\rm hard})^2 (\alphas^{\rm soft})^{1/2} = \left\{ 
  \begin{array}{lll}
    0.492^2 \times \sqrt{0.557} & = 0.181 & \mbox{\ in\ } 
            \chi_0 \to \pi^+ \pi^- \\
    0.509^2 \times \sqrt{0.566} & = 0.195 & \mbox{\ in\ } 
            \chi_2 \to \pi^+ \pi^- \\
  \end{array}
  \right. \,.
\label{alphaseff}
\end{equation}
The smaller values of the products of strong couplings have to be 
compensated for by a larger value of $f^{(8)}$. Observe that the
coupling $\alpha_s^{\mrm{soft}}$ of the constituent gluon of the 
$\chi_{\c J}$ is only slightly larger than the coupling 
of the ``hard'' gluons. This is caused by two effects: 
(i) the dynamical setting of the scales makes $\alpha_s^{\mrm{hard}}$  
larger than the naive, collinear estimate 
$\alpha_s^{\mrm{hard}}  = \alpha_s^{\mrm{hard}}(\mc)$; 
(ii) the Sudakov factor is effective enough 
to suppress the soft phase-space regions, in which 
$\alpha_s^{\mrm{soft}}$ would become large. 
The fact that this coupling is not probed at large values gives 
us confidence in our perturbative treatment of the $\chi_{\c J}$ 
cosntituent gluon. 

The question remains whether the result (\ref{foctet}) is physically
sensible or not. Therefore, we are going to estimate the probability
$P_{\c \cbar g}$ to find the $\chi_{\c J}$ in the colour-octet state
from the following examplary parametrization for the $\c \cbar \g$
wave function
\begin{eqnarray}
  \label{bsw}
  \Psi_{0}^{(8)}(z_i,{\bf k}_{\perp i}) = N z_1 z_2 z_3^2 &&\,
      \exp\left\{-2a_{\chi}^{2}\mc^2\,
                \left[ (z_3-z)^2 + (z_1-z_2)^2 \right]\right\}\nonumber\\
                &&\times \exp\left\{-a_{\chi}^2 \sum k^2_{\perp i}\right\}
\ ,
\end{eqnarray}
This ansatz combines the known asymptotic behaviour of a \da\ for a
$\q\qbar\g$ Fock state with a mass-dependent exponential and a Gaussian
$k_{\perp}$ dependence in analogy to the Bauer--Stech--Wirbel
parametrization of charmed-meson wave functions \cite{Bsw85}. The
mass-dependent exponential guarantees a pronounced peak of the \da\ at
$z_1 \simeq z_2 \simeq (1-z)/2$.  The $\delta$-function-like \da\ used
in the estimate of the colour-octet contribution appears as the
peaking approximation to this function. Since the $\c\cbar\g$ Fock
state is an $S$-wave state we assume its radius to be equal to that of
the $S$-state charmonia, namely $0.42\;{\rm fm}$ \cite{BuT81}. In this
case the transverse size parameter $a_{\chi}$ takes the value $1.23$
GeV$^{-1}$.  
The probability of the colour-octet Fock state is then found to be 
\begin{equation}
  \label{ocprob}
  P_{\c\cbar\g} = \left ( f^{(8)}/2.1\times 10^{-3}\, {\rm GeV}^{2}
                                                               \right )^2
\ .
\end{equation}
This relation is obtained with $z = 0.15$; it is however practically
independent of the exact value of $z$. We stress again that the ansatz
(\ref{bsw}) is only an example for a possible wave function. Thus, our
result of about 1/2 for $P_{\c\cbar\g}$ (from (\ref{foctet}) in
(\ref{ocprob})) is not in apparent disagreement with expectation
and we may conclude that the result (\ref{foctet}) corresponds to a
large but not unphysical value for $P_{\c\cbar\g}$.   
\begin{table}
 \begin{center} 
  \begin{tabular}{|l|c|c|} \hline
                 & $J = 0$ & $J = 2$  \\ \hline\hline
   $a_J^{(1)}$   & $23.36 + 14.67\,\imath$ & $\phantom{1}5.14 + 3.42\,\imath$ 
                 \\ \hline 
   $a_J^{(8)}$   & $33.90 + 15.89\,\imath$ & $11.60 + 4.12\,\imath$ 
                 \\ \hline\hline
   $a_J^{(1+8)}$ & $57.26 + 30.56\,\imath$ & $16.74 + 7.54\,\imath$
                 \\ \hline
  \end{tabular}
 \end{center}
 \caption[]{Contributions to $a_J$ (singlet, octet, and their sum 
      for $f^{(8)} = 1.46 \times 10^{-3} \,\mbox{GeV}^2\,$). 
  \label{tab:aJ} }
\end{table}

Using the value (\ref{foctet}) the coefficients $a_J$ add up as
shown in Table~\ref{tab:aJ}. Our results for the decay widths are presented in
Table~\ref{tab:widths}. Note that in the colour octet decay amplitude there
are terms contributing to decays into neutral pions only (see group
11 in Fig.~\ref{fig:col8diags2}). We may thus find deviations
from 1/2 for the ratio of the decay width into neutral pions over that
into charged pions. In experiment as well as in our calculation these
deviations are found to be small.  
\begin{table} 
 \begin{center}
  \begin{tabular}{|l|c|c|c|} \hline
      &\phantom{results}  &  PDG \cite{pdg96} & BES \cite{bes96} \\ \hline\hline
  $\Gamma[\chi_{\c 0} \to \pi^+ \pi^-]$ [keV] & 45.4  
     & $105 \pm 47\phantom{6}$ & $64 \pm 21\phantom{.3}$ \\ \hline
  $\Gamma[\chi_{\c 2} \to \pi^+ \pi^-]$ [keV] & 3.64
     & $3.8 \pm 2.0$ & $3.04 \pm 0.73$ \\ \hline
  $\Gamma[\chi_{\c 0} \to \pi^0\,\pi^0\,]$ [keV] &  23.5
     & $43 \pm 18$ & \\ \hline
  $\Gamma[\chi_{\c 2} \to \pi^0\,\pi^0\,]$ [keV] &  1.93
     & $2.2 \pm  0.6$  & \\ \hline
  \end{tabular}
 \end{center} 
 \caption[]{Results for the $\chi_{\c J}$ decay widths into pions
            ($f^{(8)} = 1.46 \times 10^{-3} \,\mbox{GeV}^2 \,$; $B_2=0$)
            in comparison with experimental data. The BES result for 
            $\Gamma[\chi_{\c 0} \to \pi^+ \pi^-]$ is evaluated with the
            BES result for the total width. In the other cases the PDG
            average values for the total widths are used.
  \label{tab:widths}}
\end{table}
As discussed before we perform a peaking approximation in the momentum
fraction $z$ carried by the $\chi_{\c J}$ Fock-state gluon. As in
Ref.~\cite{BKS96} we set $z = 0.15$. We have checked that our results
are only weakly dependent on the actual value of $z$ within the range
between 0.1 and 0.2 and all our conclusions remain valid. 
\begin{figure}
\[
 \psfig{figure=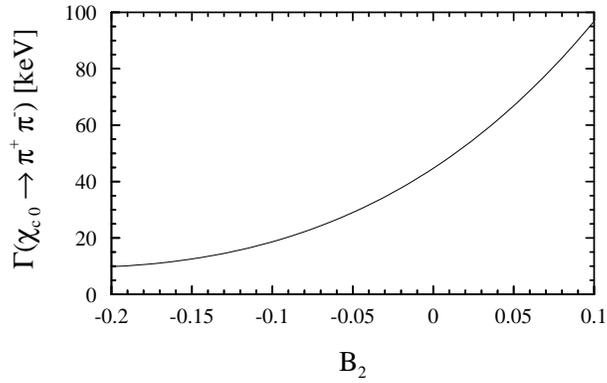,%
       bbllx=10pt,bblly=480pt,bburx=435pt,bbury=745pt,%
       height=5cm,clip=} \]
 \caption[dummy0]{Dependence of the prediction for the $\chi_{\c 0} \to
      \pi^+ \pi^-$ decay width on the expansion parameter $B_2$ of the
      pion \da.   
  \label{fig:b2dep} }
\end{figure}

In Fig.\ \ref{fig:b2dep} we sketch the dependence of the $\chi_{\c 0} \to
\pi^+ \pi^-$ decay width on the expansion parameter $B_2$ in the 
allowed $B_2$ range \cite{BKS96}. We 
dispense with a corresponding plot for $\chi_{\c 2}$ since its width is 
found to depend on $B_2$ in the same way as the width of $\chi_{\c 0}
\to \pi^+ \pi^-$. As can be seen from Fig.\ \ref{fig:b2dep} the
dependence on $B_2$ is quite strong, i.e.\ pushing $B_2$ from 
the central value zero to its upper limit $B_2 = 0.1$ doubles the width. 
If the octet decay constant
is determined by some other method, the $\chi_{\c J} \to\pi \pi$ decays
constitute a severe test of the pion \da.

We conclude from these considerations that it is possible to explain
the experimental data on $\chi_{\c J} \to \pi^+\pi^-$ within a
perturbative approach, using the asymptotic form of the pion
\da, provided colour-octet contributions are included. The numerical 
results of our new calculation within the mHSA are similar 
to our old ones \cite{BKS96}, obtained within the collinear approximation. 
The new calculation within the mHSA improves, however, the old one 
in various respects: (i) the scales of the coupling constant are determined, 
(ii) soft phase-space regions are suppressed, and (iii) all propagators 
associated with the constituent gluon are regularized through the 
non-zero transverse momenta of the pion's constituents. 
%
%
\section{$\chi_{\c J}$ decays into other light pseudoscalars}
%
\setcounter{equation}{0}
The considerations made so far for the case of the pion can 
straightforwardly be generalized to the decays of $\chi_{\c J}$ into the other
light pseudoscalar mesons, e.g.\ etas and kaons. The $\eta$ meson is a
linear combination of an SU(3) octet and singlet component
\begin{equation}
  |\,\eta\,\rangle = \cos\theta_P\:|\,\eta_8\,\rangle
               \,-\, \sin\theta_P\:|\,\eta_1\,\rangle \,.
  \label{etastate}
\end{equation}
For the wave functions of $\eta_1$ and $\eta_8$ we follow \cite{JKR96}
and make the same ansatz (\ref{pion})--(\ref{Sigma}) for it as for the 
pion (with $a_{\eta} = a_{\pi}$). The decay constants $f_{\eta_1}$
and $f_{\eta_8}$ as well as the mixing angle $\theta_P$ have been 
determined from a fit to the $\eta$--$\gamma$ and $\eta'$--$\gamma$  
transition form-factor data \cite{JKR96}:
\begin{equation}
  \hspace{-0.5cm}
  f_{\eta_8} = 145 \pm  3 {\rm\,MeV}; \hspace{1.5cm}
  f_{\eta_1} = 136 \pm 10 {\rm\,MeV}; \hspace{1.5cm}
  \theta_P = -18^{\circ} \pm 2^{\circ} 
\ .
\end{equation}
Since the strong interaction is flavour-blind, the hard scattering
amplitudes for the decays of $\chi_{\c J}$ into $\eta_1$ and $\eta_8$
pairs equal both that of the $\pi^0\pi^0$ channel (with group 11
included), i.e.~the only modification with respect to the pion case is
to replace $f_{\pi}^2$ by $\cos^2\theta_P f_{\eta_8}^2 + \sin^2\theta_P^2
f_{\eta_1}^2 = (144 \rm{\,MeV})^2$ in Eq.\ (\ref{decayzero}). In
addition we take into account the $\eta$ mass in the
phase space factor, $[1-m_{\eta}^2/\mc^2]^{1/2}$, of the decay width 
(\ref{width}), (\ref{decayzero}). Using, as in \cite{JKR96}, 
$B^{\eta}_2=0$ for the $\eta$ \da, we obtain the $\chi_{\c J}$ decay 
widths into pairs of etas as quoted in Table~\ref{tab:kaon}. With 
that asymptotic form of the $\eta$ \da, fair agreement with the data 
is achieved. On the other hand, a fit of $B^{\eta}$ to the 
$\Gamma[\chi_{\c J} \to \eta \eta]$ data yields a value of -0.036 for 
it and leads to a slightly better agreement with the data.

\begin{table}
 \begin{center}
  \begin{tabular}{|l||c|c||c|c|} \hline
      & \multicolumn{2}{|c||}{$\Gamma[\chi_{\c J} \to K^+ K^-]\;\;$ [keV]}  
    & \multicolumn{2}{|c|}{$\Gamma[\chi_{\c J} \to \eta \eta]\;\;$ [keV]} 
 \\ \hline
      & $J=0$ & $J=2$ & $J=0$ & $J=2$ \\ \hline \hline
    $B_2^K = -0.176$   & 22.4  & 1.68  & & \\ \hline
    $B_2^K = -0.100$   & 38.6  & 2.89  & & \\ \hline
    $B_2^{\eta}\; = -0.036$   & & & 24.0 & 1.91  \\ \hline
    $B_2^{\eta}\; = \phantom{-}0.000$   & & & 32.7 & 2.66  \\ \hline
    PDG \cite{pdg96} & $99 \pm 49$  & $3.0 \pm 2.2$ &
                       $35 \pm 20$  & $1.6 \pm 1.0$ 
                     \\ \hline
    BES \cite{bes96} & $52 \pm 17$  & $1.04 \pm 0.43$  & &
                     \\ \hline
  \end{tabular}
 \end{center} 
 \caption[]{Results for the $\chi_{\c J}$ decay widths into kaons
            and etas in comparison with experimental data
            ($f^{(8)} = 1.46 \times 10^{-3} \,\mbox{GeV}^2 \;$, $B_1^K=0$).
 \label{tab:kaon}}
\end{table}
The kaon state is also written in the form 
(\ref{pion}) - (\ref{Sigma}), with $f_{\pi}$ exchanged for $f_K =157.8$ 
MeV. Again in order to simplify matters, we assume $a_{K}=a_{\pi}$. 
Since the kaon consists of up and strange quarks with different masses
one may no longer expect the kaon \da\ $\phi_K$ to be symmetric under 
exchange $x \leftrightarrow 1-x$. Therefore, one has to include in the
expansion (\ref{phiex}) also the antisymmetric terms
\begin{equation}
  \phi_K(x,\mu_F) = \phi_{\as}(x)\, 
  \left[ 1 + \sum_{n=1,2,\ldots}^{\infty}\, B_n^K(\mu_0) 
   \left( \frac{\alpha_s(\mu_F) }{\alpha_s(\mu_0) } \right)^{\gamma_n}
  \, C_n^{(3/2)}(2x-1) \right]
\ ,
\label{phiKex}
\end{equation}
where e.g.~the anomalous dimension $\gamma_1$ is $ 32/81$
\cite{Brodsky80}. 
 
Information on the antisymmetric part of the kaon \da\ may be extracted
from the valence quark distribution functions of the kaon at large
$x$, where the contributions from Fock components higher than the
valence Fock state are negligible. In particular the ratio of strange
antiquarks over up quarks is sensitive to the amount of asymmetry in
the kaon \da. Unlike the cases of the nucleon and the pion, no
systematic model-independent analysis of the kaon quark distribution
functions exists so far. From calculations of the kaon's valence quark
distributions within the Nambu--Jona--Lasinio model \cite{shi93}, there
is evidence that at $x \approx 0.8$ there are about twice as many strange
antiquarks as $u$ quarks inside the $K^+$. 
Truncating (\ref{phiKex}) at $n=2$ one sees that such a ratio can be 
accommodated by a coefficient $B_1^K$ of order $-0.1$. About the
same value of $B_1^K$ is found in an instanton model \cite{DorPD}. 

Due to the symmetry of the hard scattering amplitude for $\chi_{\c J}\to
K^+ K^-$ under the simultaneous exchanges $x \leftrightarrow 1-x$ and $y
\leftrightarrow 1-y$, $B_1^K$ will enter the decay amplitude only
quadratically, i.e.\ in (\ref{decayzero}) and (\ref{Octetcoeffs}) 
the only additional terms are $d_J^{(c)} (B_1^{K})^2$. For 
$|B_1^K| \leq 0.1$ the contributions from the antisymmetric part of the 
\da\ therefore provide only tiny corrections to the decay of $\chi_{\c J}$ 
into kaons and can be neglected\footnote{This 
was already observed by the authors of \cite{CZ84}.}.
Hence, the hard scattering amplitude for $\chi_{\c J}$ decays into $K^+
K^-$ has the same form as the one for the decays into charged pions.
A inspection of Table~\ref{tab:kaon} reveals that already a small 
negative $B_2^K$ value of the order of $-0.1$ to $-0.2$ is sufficient  
to obtain fair agreement between the phase-space-corrected 
mHSA result and the available data. Admittedly, a precise value of 
$B_2^K$ cannot be determined at present because the BES data and the PDG
average values only agree within large errors; we therefore present
results for two different $B_2^K$ values in Table~\ref{tab:kaon}. 
Despite this uncertainty it seems that the kaon \da\ 
is somewhat narrower than the asymptotic one (which
is favoured for the pion). This finding appears to be plausible considering 
the comparatively large strange quark mass\footnote{
Remember that the non-relativistic case of infinitely heavy
quarks is described by a $\delta$-function-like \da.}.   
For $B_1^K=0$, $B_2^K = -0.1$ we obtain from (\ref{pionprops}) 
the following properties of the kaon's valence Fock state:
probability $0.37$, r.m.s.\ transverse momentum $381\,$MeV, 
and radius $0.44\,$fm.
%
%
\section{Comments on $\jp \rightarrow \pi^+ \pi^-$ decays}
It has long been known that there exists a problem with the pion 
form factor. The value $sF_{\pi}=0.95\pm 0.08\,$GeV$^2$ at 
$s = M_{\jp}^2$ deduced from the decay $\jp \rightarrow \pi^+\pi^-$
is much larger than the space-like one at the equivalent $Q^2$ scale
($Q^2 = -s$): $Q^2F_{\pi}=0.35\pm 0.10\,$GeV$^2$. Admittedly, the 
space-like data \cite{beb76} suffer from large systematical errors. 
The discrepancy is even more severe for $\psi(2S)\to \pi\pi$ 
($sF_{\pi}=2.67\pm 0.87\,$GeV$^2$). Unfortunately, the data on the
time-like form factor measured in $\e^+ \e^- \rightarrow \pi^+ \pi^-$ 
\cite{bol75}, suffer from low statistics and are inconclusive. While, 
within the large experimental errors, these data are in agreemnt with
the value for the time-like pion form factor as extracted from 
$\jp \rightarrow \pi\pi$, a much smaller value (close to that of
the space-like form factor) cannot be excluded.
These two differences between the time-like and space-like pion 
form factors pose a true challenge for the HSA. Within the {\em standard} 
HSA there is no difference between the form factor in the two regions
(except for interchanging $Q^2$ with $-s$). While possible
explanations for a larger value of the form factor in the time-like region
than in the space-like one within the mHSA have 
come up \cite{gou95}, the large values of the pion form factor
extracted from $\psi(nS) \rightarrow \pi\pi$ still lack 
a decent explanation. Here we want to propose a solution to this puzzle. 

The one-to-one correspondence between $sF_\pi$ and $\jp \rightarrow \pi\pi$
is based on the conventional assumption that 
in a leading-twist analysis the only contribution to
$\jp \rightarrow \pi\pi$ comes from the
electromagnetic process with one intermediate photon. The hard process is 
$O(\alpha_s^2\alphaem^2)$. Purely hadronic processes, seemingly larger 
by factors $\alpha_s^p/\alphaem^2$, are usually supposed to be suppressed by 
powers of $\LQCD/\mc$. Moreover, the rate for the $O(\alpha_s^6)$ 
contribution via three intermediate gluons is zero if the light-quark masses 
are assumed to be zero \cite{bro81}.

Recent developments \cite{Bod94} in the theory of heavy quarkonia have shown 
that the appropriate expansion parameter is not $\LQCD/\mc$, 
as is the case for light systems, but
rather the velocity $v$ of the heavy quark (or antiquark) in the bound state.  
The dominant Fock state of the $\jp$ is $|(\c\cbar)_1(^3S_1)\rangle$, 
a colour-singlet $\c\cbar$ pair in a spin-triplet $S$-wave state. 
This state decays to a pair of pions via a virtual photon 
and its rate is of order $\alphaem^2 \alpha_s^2 v^3$. 
(The factor $v^3$ arises from the squared wave function). 
Contributions from higher Fock states are suppressed by powers 
of $v$ and $\alpha_s^{\mrm{soft}}$. Factors of $v$ give the probability 
to find these soft gluons in the quarkonium, while factors of
$\alpha_s^{\mrm{soft}}$ are associated with the coupling of the soft 
gluons to the decay process.  

Corrections to the electromagnetic decay of the valence Fock state
first arise at relative order $v^4$ from the 
$|(\c\cbar)_8(^3S_1) \g\g\rangle$ Fock state and the 
$|(\c\cbar)_8(^3P_J) \g\rangle$ Fock state. In both cases the hard process
is of order $\alpha_s^4$, while the soft part scales as 
$v^3 (v^2 \alpha_s^{\mrm{soft}})^2$ ($v^3 v^2 (v^2 \alpha_s^{\mrm{soft}})$) 
for the former (latter) Fock state. (The former contains two soft 
gluons, the latter only one, but it is a $P$-wave annihilation. 
For comparison, in the conventional HSA (all gluons are purely perturbative
and the valence Fock state $\c\cbar_1(^3S_1)$ is the only relevant one), 
the diagrams corresponding to $|(\c\cbar)_8(^3S_1) \g\g\rangle$ are of 
order $\alpha_s^8$. The rate for the $O(\alpha_s^6)$ diagrams
corresponding to $|(\c\cbar)_8(^3P_J) \g\rangle$ vanishes for $m_\q 
\rightarrow 0$.) Since $v^2 \sim 0.3$ for charmonium, these
contributions from the higher Fock states can easily be as large as (or 
even larger than) the electromagnetic decay of the valence Fock
state. From these considerations it should be clear that the standard
extraction of the pion form factor from $\jp \rightarrow \pi\pi$ is
dubious.

\section{Summary}
%
In this paper we have presented a detailed analysis of $\chi_{\c J}$
decays into light pseudoscalar mesons within the framework of the
modified HSA. For sufficiently heavy-quark masses, quarkonia 
are almost non-relativistic, and corrections to the quark-potential 
model description can be organized into an expansion in $v$, the 
typical velocity of the charm quark in the meson. 
Recently it has been shown \cite{Bod94} that, for inclusive decays,  
the contribution from the higher Fock state $|\c\cbar_8(^3S_1)\g\rangle$ 
is not suppressed by $v$ with respect to the contribution from the 
valence Fock state $|\c\cbar_1(^3P_J)\rangle$ 
(i.e.\ the quark-potential-model Fock state). Here we found a similar 
situation for exclusive $\chi_{\c J}$ decays: the colour-octet 
contribution (i.e.\ the one arising from the higher Fock state 
$|\c\cbar \g\rangle$ with $\c\cbar$ in a colour-octet state) 
is not suppressed by powers of either $v$ or $1/\mc$. Hence 
the usual suppression of higher Fock states for exclusive reactions
does not hold in this case.

Owing to the $v$-expansion, the wave function of the $|\c\cbar\g\rangle$ 
Fock state of the $\chi_{\c J}$ is, to leading order in $v$, completely 
fixed apart from a single long-distance parameter, namely the 
colour-octet decay constant $f^{(8)}$. However, in our approach 
$f^{(8)}$ remains the only parameter of the colour-octet contribution to
the decay rate $\chi_{\c J} \rightarrow M\overline{M}$ ($M=\pi$, $K$, $\eta$).
We have given arguments against the use of higher 
Fock states for the pion. Therefore we conserve colour 
(an issue irrelevant for inclusive decays) by coupling the constituent 
gluon of the $|\c\cbar\g\rangle$ Fock state of the $\chi_{\c J}$ 
to the hard process. This becomes a sensible proceeding in the 
mHSA. The two reasons for this are as follows. 

First, we keep the transverse momentum of the pion's valence-quark 
momenta. In this way, no singular integrals occur. The regularization 
depends on the pion wave function, but the latter
is quite well constrained from an analysis of the photon--pion 
transition form factor in the mHSA. 
Secondly, incorporation of a Sudakov factor ensures that the coupling 
$\alpha_s$ of the constituent gluon to the hard process never becomes large.  
The coupling is evaluated dynamically and the Sudakov factor suppresses 
soft phase-space regions where the scale of $\alpha_s$ would become 
so small that a perturbative treatment were no longer justified. 

In our estimates of the decay rates $\chi_{\c J} \rightarrow \pi\pi$, 
$K K$, and $\eta\eta$, the colour-octet contribution
turned out to be of great importance in order to establish contact
with the experimental data. The value of the single long-distance 
parameter $f^{(8)}$ came out very reasonable, consistent with the 
expectation from $v$ scaling. 
Since the $|\Q\Qbar\g\rangle$ Fock state of $\chi_{\Q J}$ is neither 
$v$- nor power-suppressed, we expect also a large fraction of the
$\chi_{\b J}\to\pi\pi$ widths to originate from the $|\b\bbar\g\rangle$
state. This is indeed the case in our approach.  
Predictions for bottomonium decays are quoted in Table~4. At present
there is no data to compare with.
\begin{table}
\begin{center}
\begin{tabular}{|l|c|c|c|c|}
\hline
  $f^{(8)}\, [\, {\mrm{GeV}}^2\, ]$ &
 $\Gamma[\chi_{\b 0} \rightarrow \pi^+\pi^-]$ &
 $\Gamma[\chi_{\b 2} \rightarrow \pi^+\pi^-]$ &
 $\Gamma[\chi_{\b 0} \rightarrow \pi^0\pi^0]$ &
 $\Gamma[\chi_{\b 2} \rightarrow \pi^0\pi^0]$ 
\\ \hline
  $1.46 \times 10^{-3}$ & $20.6$ & $1.69$ & $10.5$ & $0.88$
\\ \hline
  $5 \times 10^{-3}$ & $152$ & $13.8$ & $78.2$ & $7.27$
\\ \hline
\end{tabular}
\end{center}
\caption{Decay widths in eV for $P$-wave bottomonia for two values 
of $f^{(8)}$ ($m_b=4.5\,$GeV;  $R'_P(0)=0.7\,$GeV${}^{5/2}$).}
\end{table}

In the case of exclusive $\jp$ decays, contributions from higher 
Fock states (and, hence, from colour-octet contributions) 
first start at $O(v^4)$ and are often also power-suppressed.
They can therefore be neglected in most reactions, 
e.g.\ in baryon--antibaryon decay channels, which are dominated by
the contributions from $\c\cbar$ annihilations through three
gluons. Indeed, a recent calculation \cite{bol97} along the lines
proposed here provides good results for many $B\overline {B}$
channels. We have pointed out that the situation is different 
for the $\jp$ decay into two pions. This decay is
customarily assumed to be dominated by the electromagnetic decay of 
$\c\cbar$ annihilation into a photon, since the three-gluon 
contributions cancel to zero. We have described the dominant hadronic 
decay channels arising from higher Fock states. These are likely to 
be responsible --- at least partially --- for the large difference between
the value of the pion form factor in the space-like region and its value
deduced from $\jp$ decays into pions. Our considerations may also have
consequences for the decays $\psi(nS)\to \rho\pi$ and for the process
$\gamma\gamma \to \pi\pi$ where, within the HSA, 
a substantial part of the cross-section is related 
to the pion form factor in the time-like region.

\appendix
\renewcommand{\theequation}{\Alph{section}.\arabic{equation}}
\section{Appendix: Calculation of the colour-octet contribution within the
mHSA
\label{appA}}
\setcounter{equation}{0}
%
\begin{figure}
\setlength{\unitlength}{1mm}
\begin{picture}(160,218)
 \put(  0,  0){\line(1,0){160}}
 \put(  0,  0){\line(0,1){218}}
 \put(  0,218){\line(1,0){160}}
 \put(160,  0){\line(0,1){218}}
 \put( 38,  0){\line(0,1){218}}
 \put( 89,  0){\line(0,1){218}}
 \put(0,181){\psfig{figure=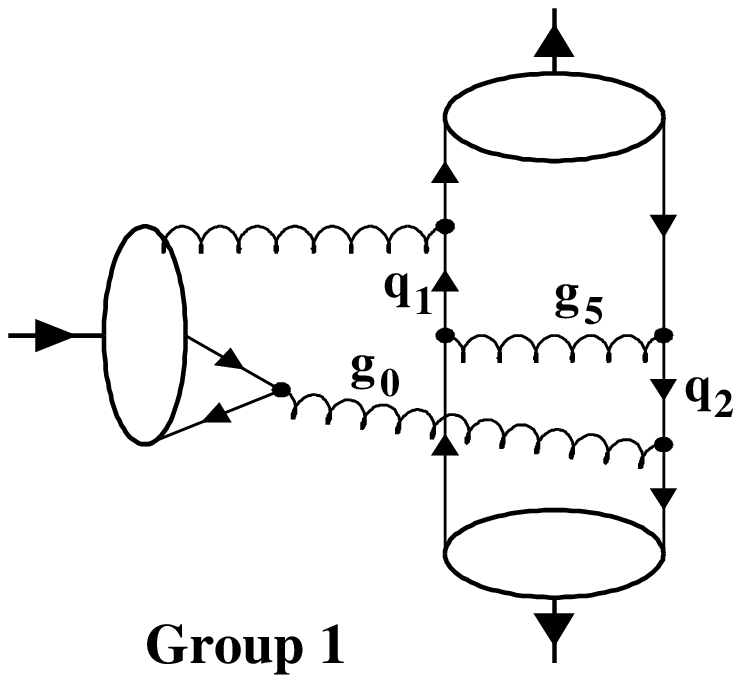,%
            bbllx=10pt,bblly=550pt,bburx=240pt,bbury=755pt,%
            clip=,width=3.5cm}}
 \put(40,195){ {\footnotesize $
  \begin{array}{rcl}
    g_0^2 & = & (1\!-\!z)^2 M^2 \\
    g_5^2 & = & (z\!-\!x_1)(z\!-\!y_1) M^2 
            - {\bf K}_{\perp}^2 \\ 
    q_1^2 & = & z(z\!-\!x_1) M^2 - {\bf k}_{\perp 1}^2 \\
    q_2^2 & = & (1\!-\!z)(y_1\!-\!z) M^2 - {\bf k}_{\perp 2}^2 \\
          &   &  \hspace{1cm} 
  \end{array} $} }
 \put( 90,198){ {\large
  $ \begin{array}{l}
     T^{(0)}_1 = \frac{\kappa_0\,M^3}{(1-z)^2}\, 
     \frac{2 z (1\!-\!z) - (x_1\!-\!z)(y_1\!-\!z)}
       {(q_1^2 + i \epsilon)(q_2^2 + i \epsilon)(g_5^2 + i \epsilon)} \\[3mm]
     T^{(2)}_1 = \frac{\kappa_2\,M^3}{(1-z)^2}\, 
     \frac{z (1\!-\!z) + (x_1\!-\!z)(y_1\!-\!z)}
       {(q_1^2 + i \epsilon)(q_2^2 + i \epsilon)(g_5^2 + i \epsilon)} \\
    \end{array} $} }
 \put(  0,180){\line(1,0){160}}
 \put(0,145){\psfig{figure=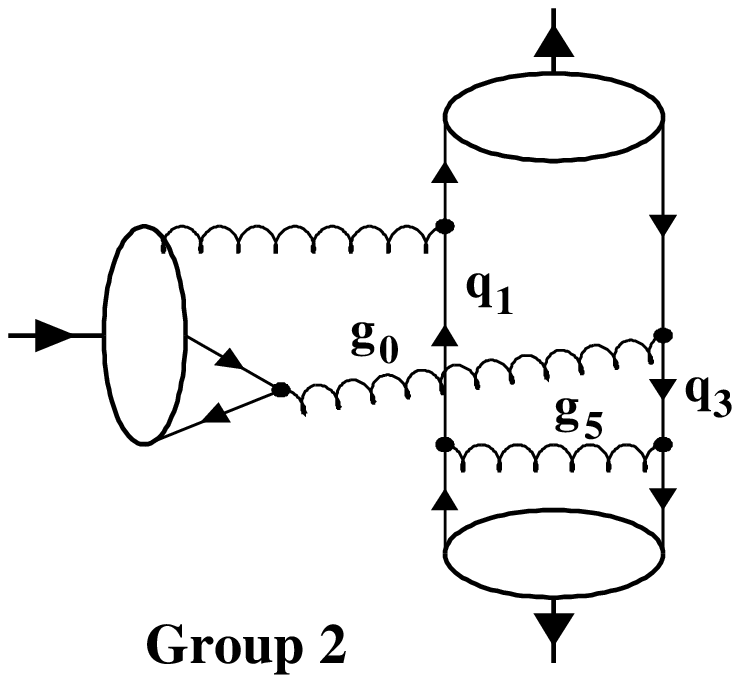,%
            bbllx=325pt,bblly=550pt,bburx=550pt,bbury=755pt,%
            clip=,width=3.5cm}}
 \put(40,159){ {\footnotesize $
  \begin{array}{rcl}
    g_0^2 & = & (1\!-\!z)^2 M^2 \\
    g_5^2 & = & (z\!-\!x_1)(z\!-\!y_1) M^2 
            - {\bf K}_{\perp}^2 \\
    q_1^2 & = & z(z\!-\!x_1) M^2 - {\bf k}_{\perp 1}^2 \\
    q_3^2 & = & (1\!-\!z)(x_1\!-\!z) M^2 - {\bf k}_{\perp 1}^2 \\
          &   &       \hspace{1cm}
  \end{array} $} }
 \put( 90,162){ {\large
  $ \begin{array}{l}
     T^{(0)}_2 = -\frac{8\,\kappa_0}{z\,(1-z)^3\,M}\, 
     \frac{1}
       {g_5^2 + i \epsilon} \\[3mm]
     T^{(2)}_2 = \frac{8\,\kappa_2}{z\,(1-z)^3\,M}\, 
     \frac{1}
       {g_5^2 + i \epsilon} \\
    \end{array} $} }
 \put(  0,143){\line(1,0){160}}
 \put(0,109){\psfig{figure=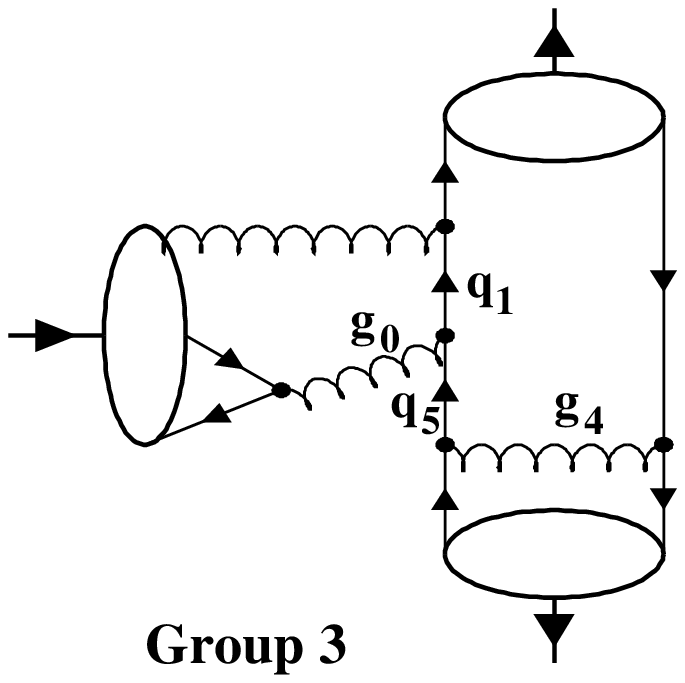,%
            bbllx=10pt,bblly=335pt,bburx=240pt,bbury=535pt,%
            clip=,width=3.5cm}}
 \put(40,123){ {\footnotesize $
  \begin{array}{rcl}
    g_0^2 & = & (1\!-\!z)^2 M^2 \\
    g_4^2 & = & x_2 y_2 M^2 
            - {\bf K}_{\perp}^2 \\
    q_1^2 & = & z (z\!-\!x_1) M^2 - {\bf k}_{\perp 1}^2 \\
    q_5^2 & = & x_2 M^2 - {\bf k}_{\perp 1}^2 \\
          &   &       \hspace{1cm}
  \end{array} $} }
 \put( 90,123){ {\large
  $ \begin{array}{r}
     T^{(0)}_3 = \frac{8\,\kappa_0\,M^3}{(1-z)^2}\,\big[ 
     \frac{2 (1\!-\!z) - (x_1\!-\!z)}
       {(q_1^2 + i \epsilon)(q_5^2 + i \epsilon)(g_4^2 + i \epsilon)} \\[-2mm] 
       {\scriptstyle + (z \leftrightarrow 1 - z)}] \\[2mm]
     T^{(2)}_3 = \frac{8\,\kappa_2\,M^3}{(1-z)^2}\,\big[ 
     \frac{(1\!-\!z) + (x_1\!-\!z)}
       {(q_1^2 + i \epsilon)(q_5^2 + i \epsilon)(g_4^2 + i \epsilon)}\\[-2mm] 
       {\scriptstyle + (z \leftrightarrow 1 - z)} ] \\
    \end{array} $} }
 \put(  0,106){\line(1,0){160}}
 \put(0, 73){\psfig{figure=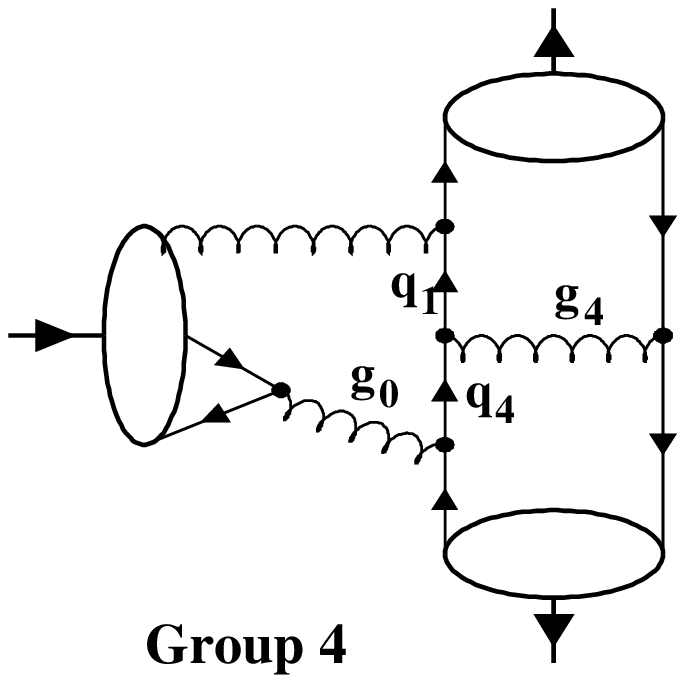,%
            bbllx=325pt,bblly=335pt,bburx=550pt,bbury=535pt,%
            clip=,width=3.5cm}}
 \put(40, 87){ {\footnotesize $
  \begin{array}{rcl}
    g_0^2 & = & (1\!-\!z)^2 M^2 \\
    g_4^2 & = & x_2 y_2 M^2 
            - {\bf K}_{\perp}^2 \\ 
    q_1^2 & = & z(z\!-\!x_1) M^2 - {\bf k}_{\perp 1}^2 \\
    q_4^2 & = & (1\!-\!z)(y_2\!-\!z) M^2 - {\bf k}_{\perp 2}^2 \\
          &   &  \hspace{1cm} 
  \end{array} $} } 
 \put( 90, 89){ {\large
  $ \begin{array}{l}
     T^{(0)}_4 = -\frac{\kappa_0}{z\,(1-z)^3\,M}\, 
     \frac{1}{g_4^2 + i \epsilon}  \\[3mm]
     T^{(2)}_4 = \frac{\kappa_2}{z\,(1-z)^3\,M}\, 
     \frac{1}{g_4^2 + i \epsilon}  \\
    \end{array} $} }
 \put(  0, 72){\line(1,0){160}}
 \put(0, 39){\psfig{figure=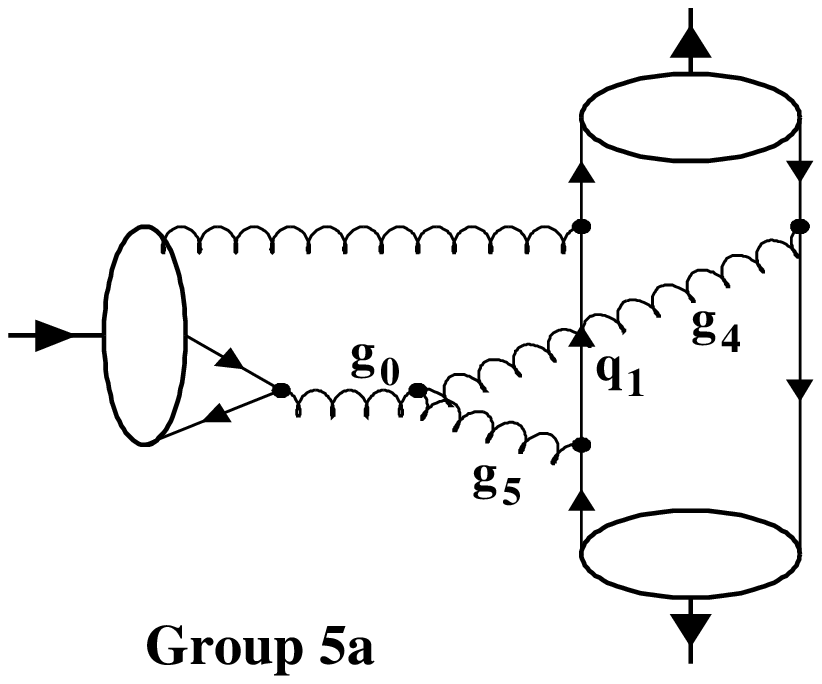,%
            bbllx=10pt,bblly=115pt,bburx=240pt,bbury=315pt,%
            clip=,width=3.2cm}}
 \put(40, 51){ {\footnotesize $
  \begin{array}{rcl}
    g_0^2 & = & (1\!-\!z)^2 M^2 \\
    g_4^2 & = & x_2 y_2 M^2 - {\bf K}_{\perp}^2 \\
    g_5^2 & = & (z\!-\!x_1)(z\!-\!y_1) M^2 
            - {\bf K}_{\perp}^2 \\ 
    q_1^2 & = & z(z\!-\!x_1) M^2 - {\bf k}_{\perp 1}^2 \\
          &   &  \hspace{1cm} 
  \end{array} $} } 
 \put( 90, 53){ {\large
  $ \begin{array}{l}
     T^{(0)}_{5a} = \frac{9\,\kappa_0\,z M^3}{(1-z)^2}\, 
     \frac{2(1\!-\!z) + (z\!-\!x_1)}
       {(q_1^2 + i \epsilon)(g_4^2 + i \epsilon)(g_5^2 + i \epsilon)} \\[3mm]
     T^{(2)}_{5a} = \frac{9\,\kappa_2\,z M^3}{2\,(1-z)^2}\, 
     \frac{2(1\!-\!z) + (z\!-\!x_1)}
       {(q_1^2 + i \epsilon)(g_4^2 + i \epsilon)(g_5^2 + i \epsilon)} \\
     \end{array} $} }
 \put(  0, 36){\line(1,0){160}}
 \put(0,  3){\psfig{figure=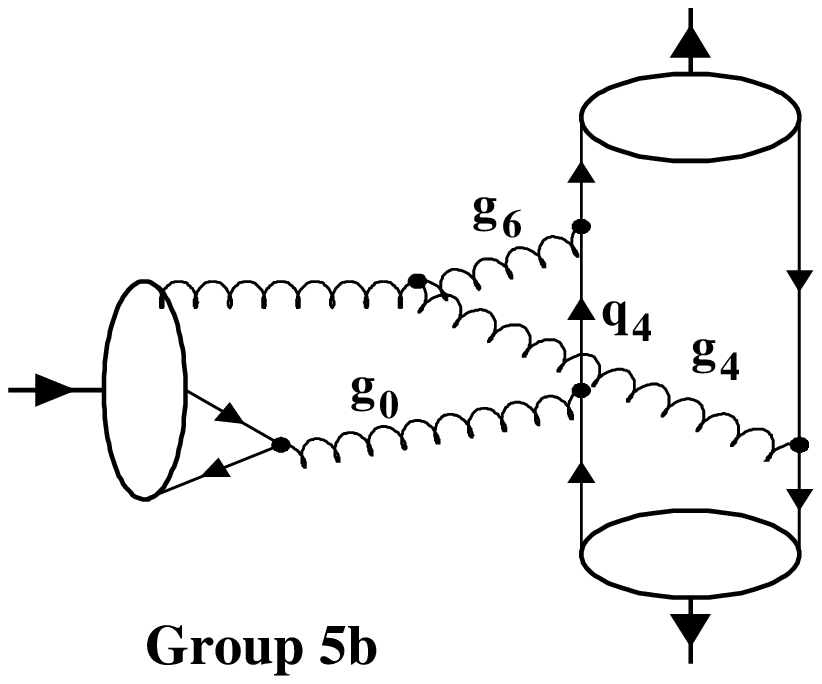,%
            bbllx=325pt,bblly=115pt,bburx=570pt,bbury=315pt,%
            clip=,width=3.5cm}}
 \put(40, 15){ {\footnotesize $
  \begin{array}{rcl}
    g_0^2 & = & (1\!-\!z)^2 M^2 \\
    g_4^2 & = & x_2 y_2 M^2 - {\bf K}_{\perp}^2 \\
    g_6^2 & = & (z\!-\!x_2)(z\!-\!y_2) M^2 
            - {\bf K}_{\perp}^2 \\ 
    q_4^2 & = & (1\!-\!z)(y_2\!-\!z) M^2 - {\bf k}_{\perp 1}^2 \\
          &   &  \hspace{1cm} 
  \end{array} $} } 
 \put( 90, 17){ {\large
  $ \begin{array}{l}
     T^{(0)}_{5b} = \frac{9\,\kappa_0\,M^3}{(1-z)}\, 
     \frac{2\,z + (y_2\!-\!z)}
       {(q_4^2 + i \epsilon)(g_4^2 + i \epsilon)(g_6^2 + i \epsilon)} \\[3mm]
     T^{(2)}_{5b} = \frac{9\,\kappa_2\,M^3}{2\,(1-z)}\, 
     \frac{2\,z + (y_2\!-\!z)}
       {(q_4^2 + i \epsilon)(g_4^2 + i \epsilon)(g_6^2 + i \epsilon)} \\
     \end{array} $} }
\end{picture}
\vspace*{5mm}
\caption[]{Overview of groups 1--5 of Feynman graphs contributing to the
colour-octet decay amplitude ($M \equiv 2\,m_c$, $x_1 \equiv x$, $x_2
\equiv 1-x$, ${\bf K}_{\perp}$ is defined in (3.7)).
\label{fig:col8diags1}}
\end{figure}
\begin{figure}
\setlength{\unitlength}{1mm}
\begin{picture}(160,218)
 \put(  0,  0){\line(1,0){160}}
 \put(  0,218){\line(1,0){160}}
 \put(  0, 32){\line(0,1){186}}
 \put(  0,  0){\line(0,1){ 31}}
 \put( 38, 32){\line(0,1){186}}
 \put( 38,  0){\line(0,1){ 31}}
 \put( 89, 32){\line(0,1){186}}
 \put( 89,  0){\line(0,1){ 31}}
 \put(160, 32){\line(0,1){186}}
 \put(160,  0){\line(0,1){ 31}}
 \put(0,181){\psfig{figure=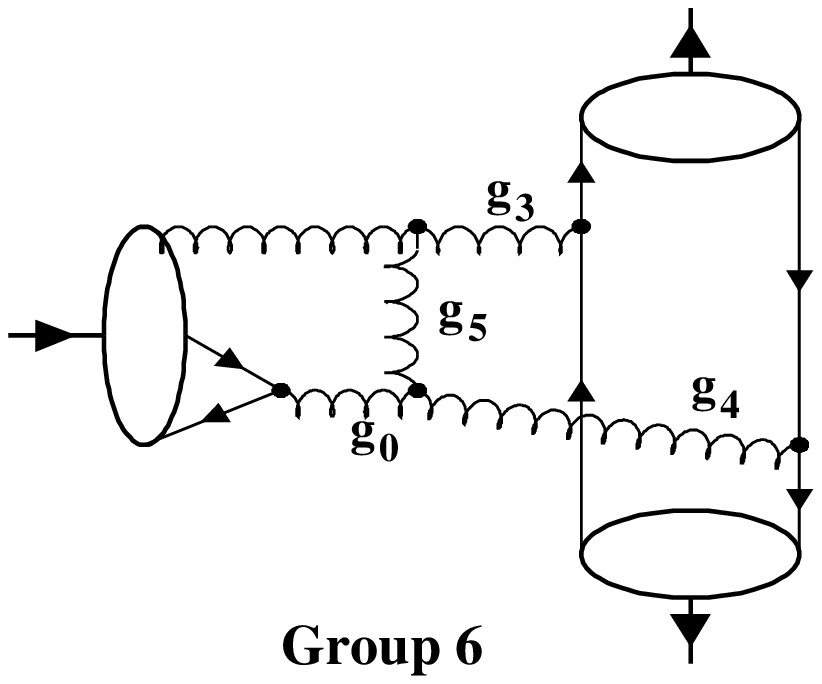,%
            bbllx=325pt,bblly=115pt,bburx=570pt,bbury=315pt,%
            clip=,width=3.5cm}}
 \put(40,195){ {\footnotesize $
  \begin{array}{rcl}
    g_0^2 & = & (1\!-\!z)^2 M^2 \\
    g_3^2 & = & x_1 y_1 M^2 - {\bf K}_{\perp}^2 \\
    g_4^2 & = & x_2 y_2 M^2 - {\bf K}_{\perp}^2 \\ 
    g_5^2 & = & (z\!-\!x_1)(z\!-\!y_1) M^2 - {\bf K}_{\perp}^2 \\
          &   &  \hspace{1cm} 
  \end{array} $} } 
 \put( 90, 198){ {\large
  $ \begin{array}{ll}
     T^{(0)}_{6} & = \frac{9\,\kappa_0\,M^3}{(1-z)^2} \,
     \frac{1 - (z\!-\!x_2)(z\!-\!y_2) + (x_2\!-\!y_2)^2}
       {(g_3^2 + i \epsilon)(g_4^2 + i \epsilon)(g_5^2 + i \epsilon)} \\[3mm]
     T^{(2)}_{6} & = \frac{9\,\kappa_2\,M^3}{2\,(1-z)^2} \\ 
   &  \frac{(z\!-\!x_1)(z\!-\!y_1) + 2\,z(1\!-\!z) + 4\,x_1 x_2 -
           2\,x_1 y_2}
       {(g_3^2 + i \epsilon)(g_4^2 + i \epsilon)(g_5^2 + i \epsilon)} \\
     \end{array} $} }
 \put(  0,180){\line(1,0){160}}
 \put(0,150){\psfig{figure=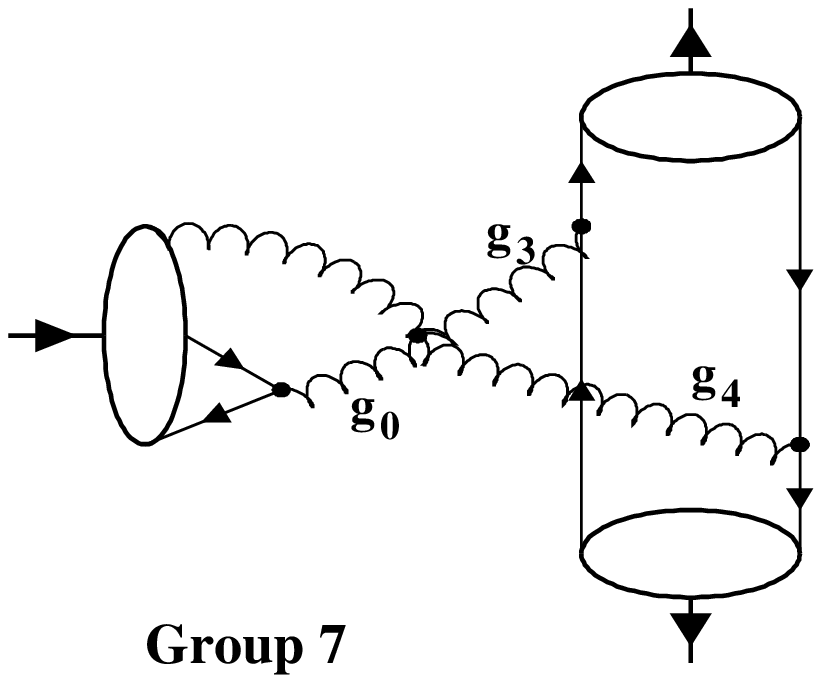,%
            bbllx=15pt,bblly=555pt,bburx=255pt,bbury=755pt,%
            clip=,width=3.5cm}}
 \put(40,162){ {\footnotesize $
  \begin{array}{rcl}
    g_0^2 & = & (1\!-\!z)^2 M^2 \\
    g_3^2 & = & x_1 y_1 M^2 - {\bf K}_{\perp}^2 \\ 
    g_4^2 & = & x_2 y_2 M^2 - {\bf K}_{\perp}^2 \\
          &   &  \hspace{1cm} 
  \end{array} $} } 
 \put( 90, 164){ {\large
  $ \begin{array}{l}
     T^{(0)}_{7} = \frac{18\,\kappa_0\,M}{(1-z)^2}\, 
     \frac{1}{(g_3^2 + i \epsilon)(g_4^2 + i \epsilon)} \\[3mm]
     T^{(2)}_{7} = -\frac{9\,\kappa_2\,M}{2\,(1-z)^2}\, 
     \frac{1}{(g_3^2 + i \epsilon)(g_4^2 + i \epsilon)} \\
     \end{array} $} }
 \put(  0,150){\line(1,0){160}}
 \put(0,113){\psfig{figure=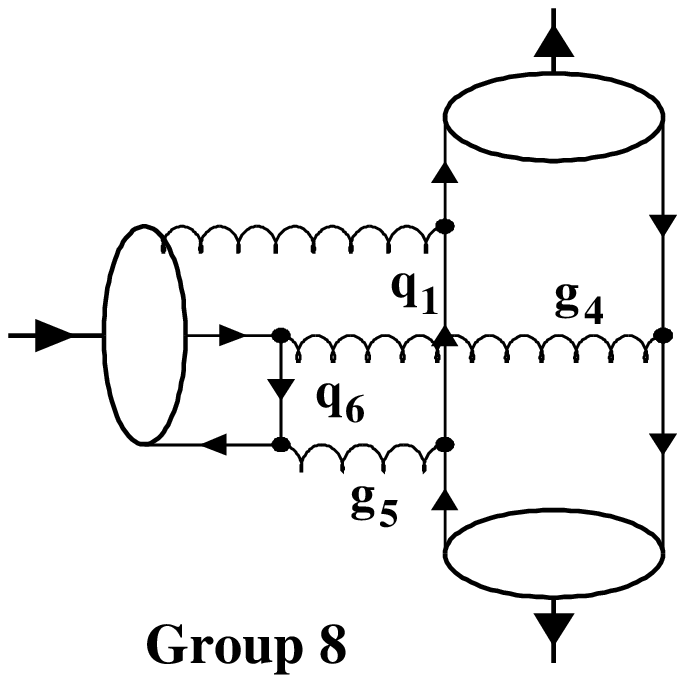,%
            bbllx=325pt,bblly=555pt,bburx=535pt,bbury=755pt,%
            clip=,width=3.5cm}}
 \put(40,127){ {\footnotesize $
  \begin{array}{rcl}
    g_4^2 & = & x_2 y_2 M^2 - {\bf K}_{\perp}^2 \\
    g_5^2 & = & (z\!-\!x_1)(z\!-\!y_1) M^2 - {\bf K}_{\perp}^2 \\
    q_1^2 & = & z (z\!-\!x_1) M^2 - {\bf k}_{\perp 1}^2 \\
    q_6^2 & = & \big( \left[x_2(z\!-\!y_1) + y_2(z\!-\!x_1)\right] \\
          & + & (1\!-\!z)^2/2 \big) M^2/2 - {\bf K}_{\perp}^2 \\ 
          &   &  \hspace{1cm} 
  \end{array} $} } 
 \put( 90, 130){ {\large
  $ \begin{array}{ll}
     T^{(0)}_{8} = & \frac{ 9\,\kappa_0\,z M^5}
      {(g_4^2 + i \epsilon)(g_5^2 + i \epsilon)(q_1^2 + i \epsilon)} \\[-1mm]
     & \frac{1 + z/2 - x_1}{q_6^2 - M^2/4 + i \epsilon} \\[2mm]
     T^{(2)}_{8} = & \frac{ 9/2\,\kappa_2\,z M^5)}
      {(g_4^2 + i \epsilon)(g_5^2 + i \epsilon)(q_1^2 + i \epsilon)} \\[-1mm]
     & \frac{1 + z/2 - x_1}{q_6^2 - M^2/4 + i \epsilon} \\
     \end{array} $} }
 \put(  0,111){\line(1,0){160}}
 \put(0, 76){\psfig{figure=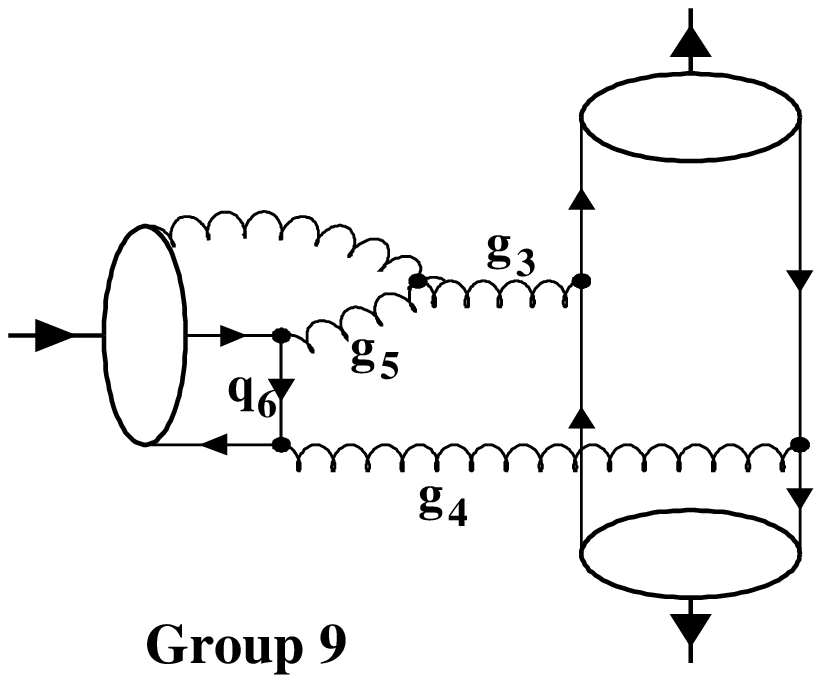,%
            bbllx=15pt,bblly=335pt,bburx=255pt,bbury=535pt,%
            clip=,width=3.5cm}}
 \put(40, 90){ {\footnotesize $
  \begin{array}{rcl}
    g_3^2 & = & x_1 y_1 M^2 - {\bf K}_{\perp}^2 \\
    g_4^2 & = & x_2 y_2 M^2 - {\bf K}_{\perp}^2 \\
    g_5^2 & = & (z\!-\!x_1)(z\!-\!y_1) M^2 - {\bf K}_{\perp}^2 \\
    q_6^2 & = & \big( \left[x_2(z\!-\!y_1) + y_2(z\!-\!x_1)\right] \\
          & + & (1\!-\!z)^2/2 \big) M^2/2 - {\bf K}_{\perp}^2 \\ 
          &   &  \hspace{1cm} 
  \end{array} $} } 
 \put( 90, 93){ {\large
  $ \begin{array}{ll}
     T^{(0)}_{9} & = {\scriptstyle 9\,\kappa_0 M^5} \\[-3mm]
         & \frac{z + z^2/2 - z x_1/2 + x_1 y_2 - y_1 (x_1-y_1)}
      {(q_6^2 - M^2/4 + i \epsilon)(g_3^2 + i \epsilon)
       (g_4^2 + i \epsilon)(g_5^2 + i \epsilon)} \\[2mm]
     T^{(2)}_{9} & = {\scriptstyle 9/2\,\kappa_2 M^5} \\[-3mm]
         & \frac{z + z^2/2 - z x_1/2 + x_1 y_2 - y_1 (x_1-y_1)}
      {(q_6^2 - M^2/4 + i \epsilon)(g_3^2 + i \epsilon)
       (g_4^2 + i \epsilon)(g_5^2 + i \epsilon)} \\
     \end{array} $} }
 \put(  0, 74){\line(1,0){160}}
 \put(0, 41){\psfig{figure=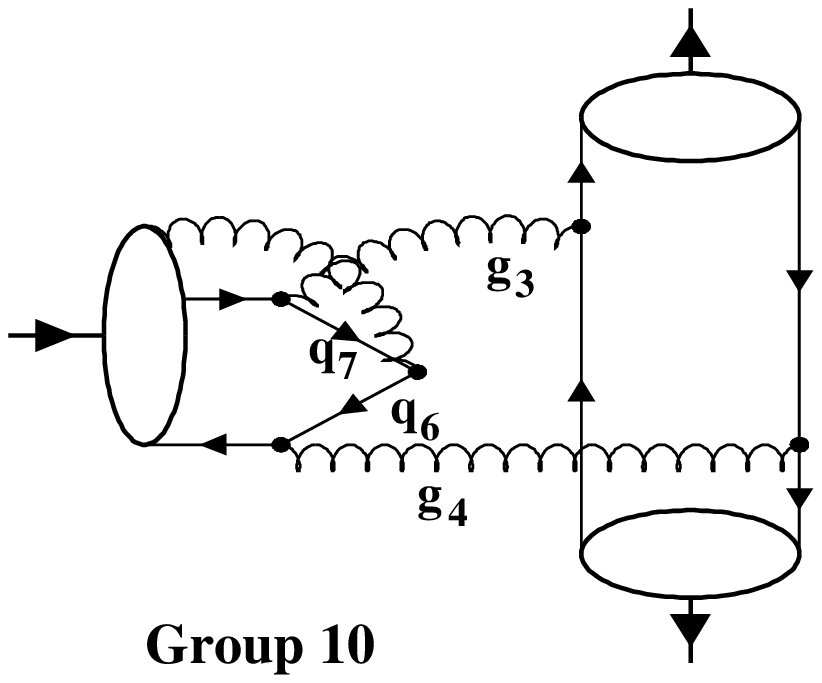,%
            bbllx=325pt,bblly=335pt,bburx=570pt,bbury=535pt,%
            clip=,width=3.5cm}}
 \put(40, 50){ {\footnotesize $
  \begin{array}{rcl}
    g_3^2 & = & x_1 y_1 M^2 - {\bf K}_{\perp}^2 \\
    g_4^2 & = & x_2 y_2 M^2 - {\bf K}_{\perp}^2 \\
    q_6^2 & = & \big( \left[x_2(z\!-\!y_1) + y_2(z\!-\!x_1)\right] \\
          & + & (1\!-\!z)^2/2 \big) M^2/2 - {\bf K}_{\perp}^2 \\ 
    q_7^2 & = & \big( \left[x_1(z\!-\!y_2) + y_1(z\!-\!x_2)\right] \\
          & + & (1\!-\!z)^2/2 \big) M^2/2 - {\bf K}_{\perp}^2 \\ 
          &   &  \hspace{1cm} 
  \end{array} $} } 
 \put( 90, 53){ {\large
  $ \begin{array}{ll}
     T^{(0)}_{10} = & -\frac{\kappa_0\,M^5}
              {4\,(g_3^2 + i \epsilon)(g_4^2 + i \epsilon)} \\
       & \frac{2\,x_1 x_2 + z^2 + 2\,z}
      {(q_6^2 - M^2/4 + i \epsilon)(q_7^2 - M^2/4 + i \epsilon)} \\[2mm]
     T^{(2)}_{10} = & \frac{\kappa_0\,M^5}
              {4\,(g_3^2 + i \epsilon)(g_4^2 + i \epsilon)} \\
       & \frac{2\,x_1 x_2 + z^2 + 2\,z}
      {(q_6^2 - M^2/4 + i \epsilon)(q_7^2 - M^2/4 + i \epsilon)} \\
     \end{array} $} }
 \put(  0, 32){\line(1,0){160}}
 \put(  0, 31){\line(1,0){160}}
 \put(0,  0){\psfig{figure=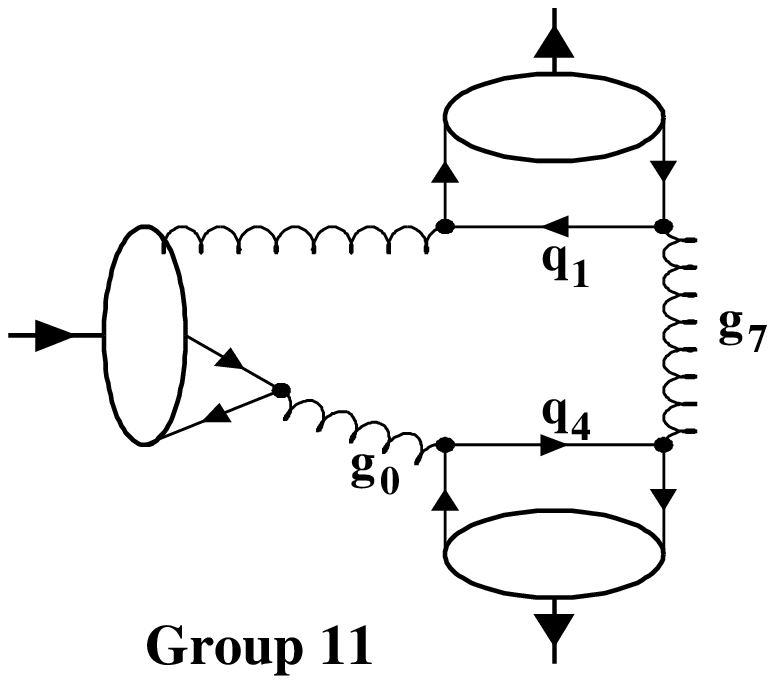,%
            bbllx=15pt,bblly=115pt,bburx=245pt,bbury=315pt,%
            clip=,width=3.5cm}}
 \put(40, 15){ {\footnotesize $
  \begin{array}{rcl}
    g_0^2 & = & (1\!-\!z)^2 M^2 \\ 
    g_7^2 & = & z (1\!-\!z) M^2 \\ 
    q_1^2 & = & z (z\!-\!x_1) M^2 - {\bf k}_{\perp 1}^2 \\
    q_4^2 & = & (1\!-\!z) (y_2\!-\!z) M^2 - {\bf k}_{\perp 2}^2 \\
          &   &  \hspace{1cm} 
  \end{array} $} } 
 \put( 90, 16){ {\large
  $ \begin{array}{l}
     T^{(0)}_{11} = -\frac{12\,\kappa_0\,M}{(1-z)^2}\, 
     \frac{1}{(q_1^2 + i \epsilon)(q_4^2 + i \epsilon)} \\[3mm]
     T^{(2)}_{11} = -\frac{9\,\kappa_2\,M}{(1-z)^2}\, 
     \frac{1}{(q_1^2 + i \epsilon)(q_4^2 + i \epsilon)} \\
     \end{array} $} }
\end{picture}
\vspace*{5mm}
\caption[]{Overview of groups 6--11 of Feynman graphs contributing to the
colour-octet decay amplitude ($M \equiv 2\,m_c$, $x_1 \equiv x$, $x_2
\equiv 1-x$, ${\bf K}_{\perp}$ is defined in (3.7)).
\label{fig:col8diags2}}
\end{figure}

%
The colour-octet decay amplitude can be written in a form similar to 
(\ref{Mb}) : 
\begin{multline}
  \label{Moct}
  M^{(8)}(\chi_{\c J}\to\pi^+\pi^-) = 
     \int_0^1 \d x \d y \int \frac{\d^2 {\bf b_1}}{4\pi}
     \frac{\d^2 {\bf b}_2}{4\pi} \,
     \hat \Psi_{\pi}^{\ast}(y,{\bf b}_2) \,
     \hat \Psi_{\pi}(x,{\bf b}_1) \\
     \times\,\sum_{i=1}^{10}\;
      \hat T_i^{(J)}(x,y,{\bf b}_1,{\bf b_2}) \,  
      \exp\left[ -S(x,y,{\bf b}_1,{\bf b}_2,t_{i1},t_{i2}) \right] 
\ .
\end{multline}
The sum runs over ten groups of graphs (see Figs.~\ref{fig:col8diags1}
and \ref{fig:col8diags2}) in the case of charged pions. For the
$\pi^0\pi^0$ final state the graphs of group 11 contribute as well. 
Each group contains a certain number of Feynman graphs, which differ only by
permutations of either the two-pion states, the light quark and
antiquark lines, or the attachment of the gluons from the charmonium
side. Note that for those Feynman graphs where the $\c\cbar$ pair
annihilates into one gluon, the momentum fractions of $\c$ and $\cbar$
do not appear explicitly and hence only the momentum fraction $z$ 
carried by the constituent gluon remains. 
For those graphs where $\c \cbar$ annihilate into two gluons 
(groups  8--10) we assume that $\c$ and $\cbar $ carry each a momentum
fraction $(1-z)/2$ of the charmonium momentum. 

In Eq.~(\ref{Moct}) the $\hat T_i^{(J)}$ denote the Fourier-transformed 
hard scattering amplitude of group $i$ for the 
colour-octet decay of the $\chi_{\c J}$ in transverse coordinate
space. The hard scattering amplitudes $T_i^{(J)}$ are given in 
Figs.\ \ref{fig:col8diags1} and \ref{fig:col8diags2}. 

In the expressions for the $T_i^{(J)}$ given in Figs.~\ref{fig:col8diags1}
and \ref{fig:col8diags2}, we absorb the combinatorial and colour
factors as well as the colour-octet $\chi_{\c J}$ decay constant and 
$\alphas$ into the factors 
\begin{eqnarray}
  \label{kappa}
  \kappa_0 &=& \frac{64\,\pi^{5/2}\,f^{(8)}}{9\sqrt{6}\,\mc^3} 
               \alphas(t_{i1}) \alphas(t_{i2}) \sqrt{\alphas(t_{i3})}\,,
       \nonumber \\ 
  \kappa_2 &=& \frac{64\,\pi^{5/2}\,f^{(8)}}{9\sqrt{2}\,\mc^3} 
               \alphas(t_{i1}) \alphas(t_{i2}) \sqrt{\alphas(t_{i3})}\,.
\end{eqnarray}
The renormalization scales $t_{ij}$ appearing in the various
amplitudes, depend on the kinematics of the specific group. The 
virtualities of the two hard internal gluons connecting the external 
lines determine the renormalization scales $t_{i1}$ and $t_{i2}$, 
respectively. In analogy to (\ref{tj}) these scales are chosen as
the maximum of $1/b_1^2$, $1/b_2^2$ and the corresponding gluon 
virtuality (at ${\bf k}_{\perp 1} = {\bf k}_{\perp 2} = {\bf 0}$).
In addition there is the constituent gluon 
from the $\chi_{\c J}$ colour-octet Fock state coupling to the hard
part. The relevant scale $t_{i3}$ is chosen as the maximum of 
$1/b_1^2$, $1/b_2^2$ and the virtuality of the adjacent internal quark
or gluon line (at ${\bf k}_{\perp 1} = {\bf k}_{\perp 2} = {\bf 0}$). 
  
We take into account the $k\trv$ dependence only in the denominators
of the $T_i^{(J)}$, i.e.\ we neglect corrections of $O (k^2_{\perp}/\mc^2)$. 
In the collinear approximation we employed in \cite{BKS96}, 
propagator singularities are touched within the range of the momentum 
fraction integrals which cannot be regularized in the usual way 
with the $\imath \epsilon$ prescription. The difficulty arises from 
the gluon propagator $g_5$ (see Figs.\ \ref{fig:col8diags1} and 
\ref{fig:col8diags2}). In \cite{BKS96} we regularized the
corresponding integrals by replacing ${\bf K}\trv$ in $g_5^2$ through 
the r.m.s.\ transverse momentum $\langle k\trv^2 \rangle^{1/2}$ of the 
quarks inside the pions. Here, in the modified mHSA all integrals
appearing in (\ref{Moct}) are either finite or can be regularized with
the $\imath \epsilon$ prescription.
In some of the $T_i^{(J)}$ there are $x$-dependent terms in the
numerators, which will cancel some of the propagator denominators if
the transverse momentum is neglected. In these cases we actually let 
these terms cancel, thus avoiding some singularities. This procedure
is equivalent to neglecting $O (k^2_{\perp}/\mc^2)$ terms. It is
applied, for instance, in group 2 through which only the dependence on
the transverse momentum difference ${\bf K}\trv$ (see (\ref{kperp})) is
left in $T_2^{(J)}$. As already encountered in the case of
the colour singlet case the corresponding Fourier-transformed hard
amplitude $\hat T_2^{(J)}$ then includes a $\delta$-function
$\delta^{(2)}(\bf b_1 - \bf b_2)$. Similar situations occur in group
4 and in parts of groups 1, 3, 5 and 8. 
  
The Fourier transform of propagators regularizable with the 
$\imath \epsilon$ prescription read 
\begin{equation}
  \int \frac{\d^2 {\bf k}\trv}{(2\pi)^2}\,
    \frac{\exp[-\imath\,{\bf k}\trv {\bf \cdot b}] }
         {s - {\bf k}\trv^2 + \imath \epsilon}
  = \left\{ \begin{array}{ll}
               -\frac{\imath}{4}\,\Ha{0}(\sqrt{s} b) & 
                           \mbox{ for}\; s > 0 \\
               -\frac{1}{2\pi}\,\BK{0}(\sqrt{-s} b) &
                           \mbox{ for}\; s < 0 
           \end{array} \right. \,.            
  \label{Hankel}
\end{equation}
Note that the inclusion of ${\bf k}\trv$ weakens the $1/s$ singularity
to a logarithmic one. In the general case of two independent
${\bf b}$-variables the integration over the relative angle $\theta$
between ${\bf b}_1$ and ${\bf b}_2$ can be analytically performed by
means Graf's theorem. For instance,  
\begin{multline}
   \int_0^{2\pi} \d \theta \,
      \Ha{0}(\sqrt{s}\,|{\bf b}_1 - {\bf b}_2|) =
   \sum_{l=-\infty}^{\infty} \int_0^{2\pi} \d \theta \,\cos(l \theta)\,
      \Ha{l}(\sqrt{s}\,\max(b_1,b_2))\,\BJ{l}(\sqrt{s}\,\min(b_1,b_2)) \\
    = 2\pi \Ha{0}(\sqrt{s}\,\max(b_1,b_2))\,\BJ{0}(\sqrt{s}\,\min(b_1,b_2))\,.
  \label{GrafTh}
\end{multline}
In most of the cases one thus finds analytical expressions for the
$\hat T_i^{(J)}$, except for group  1, where $T_1^{(J)}$ depends on ${\bf
k}_{\perp 1}$, ${\bf k}_{\perp 2}$ and ${\bf K}\trv$. Here, 
$\hat T_1^{(J)}$ includes the integral
\begin{multline}
   \int_0^{\infty} b_0 \d b_0 \,
         \Ha{0}(\sqrt{(z\!-\!x)(z\!-\!y)+\imath \epsilon}\,M b_0) \\
 \times\,\Ha{0}(\sqrt{z(z\!-\!x)+\imath \epsilon}\,M \max(b_1,b_0))  
       \,\Ha{0}(\sqrt{(1\!-\!z)(z\!-\!y)+\imath \epsilon}\,M
           \max(b_2,b_0)) \\
 \times\,\BJ{0}(\sqrt{z(z\!-\!x)+\imath \epsilon}\,M \min(b_1,b_0)) \, 
       \,\BJ{0}(\sqrt{(1\!-\!z)(z\!-\!y)+\imath \epsilon}\,M
           \min(b_2,b_0))\,. 
   \label{Hankelint}
\end{multline}
To proceed we split the range of integration at some large value
$\bar{b}_0$ of $b_0$. Actually $\bar{b}_0$ is chosen in such a way
that all arguments of the Bessel functions $\Ha{0}$ and $\BJ{0}$ are
greater than 2 for $\bar{b}_0 \geq b_0$, through which all Bessel
functions can be replaced by their asymptotic expressions and
(\ref{Hankelint}) is solved. For the remainder of the integration
region ($\bar{b}_0 \le b_0$) the integral is evaluated numerically.
%

%
\end{document}